\documentstyle[preprint,aps]{revtex}
\begin{document}
\draft
\title{ALGEBRAIC ISOMORPHISM IN TWO-DIMENSIONAL ANOMALOUS GAUGE THEORIES
\footnote[0]{38 manuscript pages}}
\author{ \sc{ C. G. Carvalhaes$^{1,3}$, L. V. 
Belvedere$^1$, H. Boschi 
Filho$ ^{2}$ },\\
\sc{C. P. Natividade$^{1}$}\\
$1$\small{\it Instituto de F\'{\i}sica - Universidade Federal Fluminense}\\
\small{\it Av. Litor\^anea, S/N, Boa Viagem, Niter\'oi, CEP.24210-340, Rio de Janeiro - 
Brasil}\\
$2$\small{\it Instituto de F\'{\i}sica - Universidade Federal
do Rio de Janeiro}\\
 \small{\it Ilha do Fund\~ao, CEP.21945, Rio de Janeiro, Brasil }\\
$3$\small{\it Instituto de Matem\'atica e Estat\'{\i}stica - Universidade 
do Estado do Rio de Janeiro}\\
\small{\it Rua S\~ao Francisco Xavier, 524, Maracan\~a,
CEP. 20559 900, Rio de Janeiro, Brasil}}
\maketitle

\begin{abstract}

The operator solution of the anomalous chiral Schwinger model is discussed 
on the basis of the general principles of Wightman field theory. Some basic
structural properties of the model are analyzed taking a careful
control on the Hilbert space associated with the Wightman functions.
The isomorphism between gauge noninvariant 
and gauge invariant descriptions of the anomalous theory is
established in terms of the corresponding field algebras. We show 
that $ (i) $ the $ \Theta $-vacuum representation and $ (ii) $ the suggested 
equivalence of vector Schwinger model and chiral 
Schwinger model 
cannot be established in terms of the intrinsic 
field algebra.

\end{abstract}

\newpage
\noindent Isomorphism in Anomalous Theories\\
Luiz Victorio Belvedere\\
Instituto de F\'\i sica\\
Universidade Federal Fluminense\\
CEP: 24210-340 Niter\'oi - RJ - Brazil\\
Fax: 55 021 620-6735\\
E-mail: belve@if.uff.br

\newpage

\section{Introduction}

The chiral Schwinger model (CSM) [1] has been intensively studied [2]
since it provides a very instructive theoretical laboratory for studying the
possibility of constructing a consistent quantum gauge theory exhibiting
anomalous breaking of a local gauge symmetry.

Following the proposal of Faddeev and Satashvili [3], in a quantum theory 
where gauge invariance is spoiled by anomalies, gauge
invariance can be restored with the introduction of {\it ad hoc} extra degrees
of freedom into the theory by adding to the original Lagrangian a 
Wess-Zumino (WZ) term [4]. 

However, several authors [5-6]  realized that the WZ term does not
need to be introduced {\it ad hoc} in order to naturally embed an
anomalous theory into a gauge theory. Using the path integral
formalism and the Faddeev-Popov prescription, the WZ action emerges as 
being related to the Jacobian of the change in the fermionic measure due to a chiral gauge 
transformation. From the operator point of view the gauge invariant 
($GI$) formulation is obtained by performing an operator-valued gauge
transformation on the gauge noninvariant ($GNI$) formulation
[7]. 

Using the path-integral formalism and the Hamiltonian formulation, it was
shown in Ref.[6] that {\it anomalous
chiral gauge theories exhibit a peculiar feature which allows two isomorphic 
formulations: the $GNI$ and
the $GI$ formulations}. From the path-integral point of view, the isomorphism is 
established between the Wightman
functions that are representations of the local field algebra
generated from the intrinsic irreducible set of field operators defining the $GNI$
formulation, and the corresponding Wightman functions of the gauge transformed 
field operators defining the $GI$ formulation of the model. Using the Hamiltonian
formalism it was shown that for the $GI$ formulation of the CSM one can
construct a set of gauge invariant operators whose equal-time
commutator algebra is isomorphic to the equal-time commutator algebra
of the phase space variables of the $GNI$ formulation of the model.
This implies that the constraints and the canonical Hamiltonian of $GI$
description map into the corresponding quantities of the $GNI$ one. As was
stressed in Ref. [6], {\it the isomorphism between the two formulations
means that the $GNI$ formulation describes the gauge invariant sector of
the enlarged $GI$ formulation}.

However, in spite of the existence of a sizable number of works on this
subject, some questions related with basic structural properties of
the anomalous chiral model have not been fully appreciated and clarified in the 
preceding literature, such as : ($i$) the implementation at the operator 
level of the isomorphism between the $GNI$ and $GI$ formulations of the CSM, as 
proposed in Ref.[6] in the path-integral and Hamiltonian formalisms; ($ii$) the 
need for a $\Theta$-vacuum parametrization in the model, as claimed in 
Ref.[9] and ($iii$) the equivalence of the vector Schwinger model (VSM) and
the $GI$ formulation of CSM defined for the regularization dependent
parameter $a = 2$ as suggested in Refs.[9,10]. It is the purpose of this 
paper to discuss the fine mathematical aspects of these questions, as
well as, to acquire a more detailed understanding of the physical grounds
of the gauge invariance implied by the addition of the WZ piece into
the anomalous theory.

We shall discuss the operator solution of the anomalous CSM on the
basis of the general principles of Wightman field theory. The guideline of
our approach of the anomalous theory will be inspired in the
rigorous treatment given in Ref.[11] for the VSM. In this approach we
only use the field algebra generated from the intrinsic irreducible
set of field operators of the theory [16] in order to exert a careful control 
on the Hilbert space associated with the Wightman functions that define the 
theory. As we shall see, by relaxing this rigorous control on the 
construction of the 
Hilbert space of the anomalous theory some misleading conclusions can
arise, as for example the existence of an infinite number of states which are 
degenerated in energy with the vacuum state [9] and the equivalence of CSM and
VSM, as proposed in Refs. [9,10]. In order to make clear that the above
conclusions rest on the incorrect choice of the observables
representing the fermionic content of the model, we shall adopt the
same bosonization scheme employed in Ref.[9].

Using the standard Mandelstam bosonization scheme the CSM is considered 
in Ref.[12]. The operator solution used introduces a minimal
Bose field algebra ({\it minimal bosonization scheme}) such that no spurious degrees of freedom
appears, and the Maxwell equation displaying the dynamical breakdown of
gauge invariance holds in the strong form and is satisfied as an operator 
identity. 

In Refs.[8,9] the vector and chiral Schwinger models are considered through
a quantization procedure that takes into account the fact that the
corresponding bosonized Lagrangian densities contain higher derivative terms 
and derivative interactions. This bosonization scheme for
higher-derivative field 
theories ({\it enlarged bosonization scheme}) introduces degrees of
freedom quantized with negative metric that generate spurious
operators. This approach resembles the local gauge formulation of
quantum field theory in which we have locality at the expense of
weakening the local Gauss' law.  The Gauss' law is modified by the
introduction of a longitudinal current and the Hilbert space
realization of the theory contains unphysical states. In the vector case 
the solution obtained in Refs.[8,9] coincides
with the Lowenstein-Swieca solution [13]. Nevertheless, due to the absence
of gauge invariance, in the anomalous CSM the additional 
gauge degree of freedom is not eliminated and the solution exhibits an enlarged
Bose field algebra that contains redundant spurious operators. In spite of the 
existence of redundant degrees of 
freedom in the solution given in Refs.[8,9], by taking into 
account the Hilbert space that is the representation of
the Wightman functions defining the model, we show that the physical content 
of the solutions given in Refs.[8,9] and by the minimal bosonization
scheme used in Ref.[12] is the same, since 
they are related by a superfluous spurious
phase operator. Since in the CSM the cluster decomposition property is 
not violated for Wightman functions that are  representations of the intrinsic 
field algebra that defines the model\footnote{ {\it In agreement with the
results of Refs.[8,12].}}, the $ \Theta $-vacuum 
parametrization suggested by the authors of Ref.[9] becomes completely
unnecessary. Of course, the suggested equivalence of the CSM
defined for $ a = 2 $ and the VSM cannot be established if we consider
the Hilbert space in which the intrinsic field algebra 
of the model is represented. The equivalence proposed in Ref.[9] is a
consequence of an incorrect decomposition of the closure of the Hilbert space
that implies the choice of a field operator that does not belong to the
intrinsic local field algebra to represent the
fermionic content of the model. Within the approach based on the
intrinsic field algebra, the existence of $ \Theta $-vacuum and the
equivalence with the VSM suggested in Refs.[9,10] cannot be regarded
as being structural properties of the CSM since they are
dependent on the use of a redundant algebra, rather than on the field
algebra which defines the model.

The paper is organized as follows. In
section $ 2 $ the $GNI$ formulation of the model is discussed and we show
that the cluster decomposition property is not violated for the Wightman
functions that are representations of the field algebra generated
from the irreducible set of field operators defining the model. The
case $ a = 2 $ is considered and we show that the factorization of
the completion of states performed in Refs.[9,10] leads to some
improper conclusions about basic structural properties of the model. In
section $ 3 $ we consider the $GI$ formulation of the model and the
isomorphism between the intrinsic operator field algebra defining the $GI$ 
formulation and the intrinsic field algebra defining the $GNI$
formulation is established. The role played by the WZ field in the 
implementability of extended local gauge transformations is also discussed.

\section{Gauge Noninvariant Formulation} 

The CSM is a two-dimensional field theory defined from the classical Jackiw-Rajaraman (JR) Lagrangian
density\,\footnote{ \it {The 
conventions used are: \,$ \psi \,=\,
(\,\psi _{_{\!\!\ell}} \,,\,\psi _{_{\!\!r}})^{\,T} $\,\,,\,\,$ \epsilon ^{\,0\,1}\,=\,g^{\,0\,0}\,
=\,-\,g^{\,1\,1}\,=\,1 $\,\,,\,\,
$ \tilde \partial ^{\,\mu }\,\equiv \,\epsilon ^{\,\mu \,\nu}\,
\partial _{\nu} $ \,\,,
\newline 
\centerline{$ \gamma ^{\,5}\,=\,-\,\gamma ^{\,0}
\gamma ^{\,1}\,\,,\,\,\gamma ^{\,\mu }\gamma ^{\,5}\,=\,\epsilon ^{\,\mu \,\nu}
 \,\gamma _{\nu } $\,\,,} 
\newline
\centerline{$ \gamma ^{\,0}\,=\,\pmatrix {0&1 \cr 1&0} $\,\,,\,\, 
$ \gamma ^{\,1}\,=\,\pmatrix {0&1 \cr -1&0} $ \,\,,\,\, $ \gamma ^{\,5}\,=\,
\pmatrix {1&0 \cr 0&{-1}}\,\,. $}}} [1]
$$ {\cal L}_{_{JR}}\,=\,-\,\frac{1}{4}\,
(\,{\cal F}_{\mu \nu }\,)^{\,2}\,
+\,i\,\bar \psi \,\partial _{\mu }\,\gamma ^\mu \,\psi\,- \,\,g\,
{\cal A}_{\mu }\,
\bar \psi \,{\gamma }^{\mu }\,P_{+}\,\psi \,\,.
\eqno (2.1) $$
Here \,$ {\cal F}_{\mu \,\nu } $ \,denotes, as usual, the field-strength tensor and $ P_{+}\,=\,
(1+{\gamma }^5) $ 
projects out the left-moving fermion. The classical Lagrangian density (2.1) exhibits
invariance under the local gauge transformation
$$ \psi \,\longrightarrow \,e^{\,i\,P_{+}\,\Lambda (x)}\,\psi \,\,,\eqno (2.2a)$$
$$ {\cal A}_{\mu }\,\longrightarrow\,{\cal A}_{\mu
}\,+\,\frac{1}{g}\,\partial _{\mu }\,\Lambda (x)\,\,. \eqno (2.2b)$$

At the quantum level the chiral anomaly spoils the gauge invariance
and the model is described by the gauge noninvariant effective Lagrangian [7],
$$  {\cal L}_{_{_{GNI}}}\,=\,-\,\frac{1}{4}\,(\,{\cal F}_{\mu \nu }\,)^{\,2}\,
+\,i\,\bar \psi \,\partial _{\mu }\,\gamma ^\mu \,\psi\,+\,\frac{g}{\sqrt \pi}\,
{\cal K}_{_{\!\ell}}^{\,\mu}\,{\cal A}_{\mu }\,+\,\frac{a g^{\,2}}{2\pi}
\,({\cal A}_{\mu })^{\,2}\,\,,
\eqno (2.3)$$
where $a$ is the JR parameter that characterizes the ambiguity in
the quantization of the model and \,$ {\cal K}_{_{\!\ell}}^{\,\mu } $\,\,is the 
left-current defined by the point-splitting regularization, independent of 
the JR parameter [7]
$${\cal K}_{_{\!\ell}}^{\,\mu }\,=\,2\,\sqrt \pi \,\lim_{\varepsilon
\rightarrow 0}\,\Bigl [\, \bar \psi (x + \varepsilon)\,{\gamma }^{\mu }\,P_{+}\,
\psi (x)\,-\,V.E.V.\,\Bigr ]\,\Bigl [\,1\,+\,i e \sqrt \pi
\,P_+\,(\,g_{\mu \nu}\,-\,\epsilon _{\mu \nu}\,)\,\varepsilon^\mu
{\cal A}^\nu (x)\,\Bigr ] \,\,. \eqno (2.4) $$

The bosonized effective $GNI$ Lagrangian density 
is [1]
$$ {\cal L}_{_{_{GNI}}}\,= \,-\,\frac{1}{4}\,(\,{\cal F}_{\mu \nu }\,)^{\,2}\,
+\,\frac{1}{2}(\partial ^{\,\mu}\phi )^{\,2}\,+\,\frac{g}{\sqrt \pi}\,
(\partial ^{\,\mu } \phi\,+\,\tilde \partial ^{\,\mu } \phi ){\cal A}_{\mu }\,
+\,\frac{a g^{\,2}}{2\pi}({\cal A}_{\mu })^{\,2}\,\,,\eqno (2.5) $$
and the resulting equations of motion are
$$ \Box \,\phi \,+\,\frac{g}{\sqrt \pi}\,(\,\partial ^\nu\,+\,\tilde
\partial ^\nu\,)\,{\cal A}_\nu\,=\,0\,\,, \eqno (2.6a) $$
$$ \partial _{\mu}\,{\cal F}^{\,\mu \nu}\,=\,J^{\,\nu}\,=
\,-\,\frac{g}{\sqrt \pi}\,(\,\partial ^{\,\nu}\,
+\,\tilde \partial ^{\,\nu }\,)\,\phi\,-\,a\,\frac{g^2}{\pi }\,
{\cal  A}^{\,\nu }\,. \eqno (2.6b) $$

Introducing a local gauge decomposition for the gauge field 
$$ {\cal A}^{\,\mu}\,=\,\tilde \partial ^{\,\mu }\,\chi\,+\,\partial ^{\,\mu}\,
\lambda \,\,,\eqno (2.7) $$
the field-strength tensor is given by
$$ {\cal F}^{\,\mu\,\nu}\,=\,-\,\epsilon ^{\,\mu \,\nu}\,\Box \,\chi \,\,.
\eqno (2.8) $$
The Maxwell Lagrangian can be written as
$$ {\cal L}_{_M}\,=\,-\,\frac{1}{4}\,({\cal F}^{\,\mu\,\nu})^{\,2}\,=\,
\frac{1}{2}\,\Box\,\chi \,\Box \,\chi \,\,,\eqno (2.9) $$
and the bosonized Lagrangian density (2.5) can be treated as a 
higher-derivative field theory [8,9,14].

For our purposes it is therefore convenient to start with the
{\it enlarged bosonization scheme} used in 
Refs.[8,9,14] and introduce the field transformations ($ a \neq 1 $)\,\footnote{ \it{For the
benefit of the reader we shall use the same notation adopted in Refs.[8,9,14].}} 
$$ \phi \,=\,\phi ^{\,\prime}\,+\,\frac{g}{\sqrt \pi}\,\Bigl ( \chi\,-\,
\lambda \,\Bigr ) \,\,,\eqno (2.10a) $$
$$ \chi _{_{{}_1}}\,=\,\frac{1}{m_a}\,\Box \,\chi \,\,, \eqno (2.10b) $$
$$ \chi _{_{{}_2}}\,=\,\frac{1}{m_a}\,(\,\Box \,+\,m_a^{\,2}\,)\,\chi \,\,,
\eqno (2.10c) $$
$$ \lambda \,=\,\lambda^{\prime}\,-\,\frac{1}{a-1}\,\chi \,\,,\eqno (2.10d) $$
with $ m_a\,=\,g\,a\,/\,\sqrt {\pi \,(a\,-\,1\,)} $. The Lagrangian density
(2.5) of the $GNI$ formulation is then reduced to a local one
\footnote{ \it As stressed in Ref.[8], these higher-derivative 
field theory transformations are consistent with the Dirac brackets of the 
constrained theory obtained in Refs.[18].} [8,9]
$$ {\cal L}_{_{_{GNI}}}\,=\,\frac{1}{2}\,(\,\partial _{\mu}\,\phi ^{\,\prime}\,)^{\,2}\,-\,\frac{1}{2}\,
(\,\partial _{\mu}\,\chi _{_2}\,)^{\,2}\,+\,\frac{g^{\,2}}{2\pi}(a\,-\,1)\,(\,\partial _{\mu}
\,\lambda^{\,\prime}\,)^{\,2}\,+\,\frac{1}{2}\,(\,\partial _{\mu}\,\chi_{_1}\,)^{\,2}\,
-\,\frac{1}{2}\,m_a^{\,2}\,\chi _{_1}^{\,2}\,\,. \eqno (2.11) $$
The field $ \chi _{_2} $ is a free and massless field quantized with negative
metric, and $ \lambda ^{\,\prime} $ is a non-canonical free massless field. In
the VSM the gauge invariance ensures that the 
field $ \lambda {^\prime} $ is a pure gauge excitation and does not 
appears in the bosonized Lagrangian density [8,14]. However, {\it in
the anomalous chiral model the additional degree of
freedom $ \lambda ^{\prime} $ is a dynamical field and, as we shall see,
its presence ensures the existence of fermions in the asymptotic
states}, implying that the screening and confinement aspects exhibited
by the CSM differ from those of the VSM \footnote{ \it As we shall see
the non-trivial anomalous nature 
of the field $ \lambda ^\prime $ ensures the dynamics for the WZ field.}. This behavior occurs for all
values of the regularization dependent parameter $ a > 1 $.

Let us consider the equation of motion (2.6b). In terms of the new fields 
(2.10), we can rewrite the vector current as
$$ J^{\mu }\,=\,m_a\,\,\tilde \partial ^{\mu }\,\chi _{_1}\,+\,L^{\,\mu }\,\,,
\eqno (2.12) $$
where $ L^{\,\mu } $ is a longitudinal 
current ($ \Box \,L^{\,\mu }\,=\,0 $) given in terms of derivatives of the 
massless fields by
$$ L^{\,\mu }\,=\,-\,\frac{g}{\sqrt \pi }\,\Biggl \{\,(\,\partial ^{\,\mu }\,
+\,\tilde \partial ^{\,\mu }\,)\,\phi ^{\,\prime }\,+
\,\frac{a}{\sqrt {a\,-\,1}}\,\tilde \partial ^{\,\mu }\,\chi _{_2} \,
+\,\frac{g}{\sqrt \pi}\Bigr [ (a-1)\partial
^{\,\mu}\lambda^{\prime}\,-\,\tilde
\partial^{\,\mu}\lambda^{\prime}\,\Bigr ]\,\Biggr \}\,\,. \eqno (2.13) $$
The first term in Eq.(2.13) represents the free left-moving Fermion 
current $ j^{\mu}_{_{\ell}} $. The negative metric quantization of the field $ \chi _{_2} $ ensures 
that 
$$ \bigl [\,L^{\,\mu }(x)\,,\,L^{\,\nu }(y)\,\bigr ]\,=\,0\,\,
\,,\forall \,x,y\,\,, \eqno (2.14) $$
and $ L^{\,\mu } $ generates, from the vacuum, zero norm states
$$ \langle \,L^*_{\mu } \,L_{\nu }\,\rangle_o \,\equiv \, \langle \,
L_{\mu } \Psi_o\,,\,L_{\nu } \Psi _o
\,\rangle\,=\,0\,\,, \eqno (2.15) $$
where $ \Psi_o $ is the vacuum vector.

For further convenience we introduce the dual field [8,9]
$$ \tilde \phi \,=\,\tilde \phi^{\prime}\,-\,\frac{g}{\sqrt \pi}\,(\chi
- \lambda )\,\,, \eqno (2.16) $$
which enables the enlargement of the algebra of the Bose fields by the
introduction of two independent right and left mover fields
$$ \phi _{_{\!r}} = \frac{1}{2} (\phi + \tilde \phi ) = 
\phi _{_{\!r}}^{\,\prime} \,\,,
 \eqno (2.17a)$$
$$ \phi _{_{\!\ell}} = \frac{1}{2} (\phi - \tilde \phi ) =\,\phi _{_{\!\ell}}^{\,\prime}\,+ \,
\frac{1}{\sqrt {a - 1}}\,( \chi _{_2} - \chi _{_1} )\,-\,\frac{g}{\sqrt \pi }\,
\lambda^{\prime}\,\,,   \eqno (2.17b)$$
$$ L_{_{\!r}} = \frac{1}{2}\,( L + \tilde L)\,=\, \frac{g}{\sqrt \pi}\,a\,
 \lambda^{\,\prime}_{_{r}}\,-\,\frac{a}{\sqrt {a - 1}}\,\chi _{_{2 r}}  \,\,,
\eqno (2.17c)$$
$$ L_{_{\!\ell}} = \frac{1}{2} (L - \tilde L)\,=\, 2\,\phi ^{\,\prime }_{_{\!\ell}}\,+\,
\frac{g}{\sqrt \pi}(a - 2)\,\lambda^{\prime}_{_{\ell}}\,+\,
\frac{a}{\sqrt {a-1}}\,\chi_{_{2\ell}}\,\,, \eqno (2.17d) $$
where $ L $ is the potential for the longitudinal current (2.13)
$$ L^{\,\mu}\, = \,-\, \frac{g}{\sqrt \pi }\, \partial ^{\,\mu } L \,= \,
\varepsilon^{\,\mu \nu} L^{\,5}_{\mu}\,=\,\frac{g}{\sqrt \pi }\,
\tilde \partial ^{\,\mu } \tilde L  \,\,.
\eqno (2.18)$$

The operator solution of the equations of motion is constructed from
the set of Bose 
fields $ \{\,\chi_{_1}, \chi _{_2}, \phi ^\prime, \lambda ^\prime \}
$, the so-called ``{\it building blocks}'' [11], and is given by
$$ {\cal A}^{\,\mu}(x)\,=\,\frac{\sqrt \pi}{g a \sqrt{a-1}}\,\Bigl
\{\,(a - 1) \tilde \partial^{\,\mu}\,( \chi_{_2}(x)\,-\,\chi_{_1}(x)\,)\,-\,
\partial^{\,\mu}\,( \chi_{_2}(x)\,- \,\chi_{_1}(x)\,)\,\Bigr \}\,+\,\partial^{\,\mu}
\lambda^{\prime}(x)\,\,, \eqno(2.19a) $$
$$ \psi_{_{\!\!r}}^0(x)\,=\,\Bigl (\frac{\mu_o}{2 \pi} \Bigr
)^{1/2}\,:\,e^{\,2i\sqrt \pi \,\phi_{_{r}}(x)}\,:\,=\,
\Bigl (\frac{\mu_o}{2 \pi} \Bigr)^{1/2}\,:\,e^{\,2i\sqrt \pi \,
\phi_{_{r}}^{\prime}(x)}\,:\,\,\,, \eqno(2.19b) $$
$$ \psi_{_{\!\!\ell}}(x)\,=\,\Bigl (\frac{\mu_o}{2 \pi} \Bigr
)^{1/2}\,:\,e^{\,2i\sqrt \pi
\,\phi_{_{\ell}}(x)}\,:\,=\,:\,e^{\,-2i\sqrt{\pi\,/\,(a - 1)}\,\chi_{_1}}(x)\,:\,
\psi_{_{\ell}}^{\,0}(x)\,\Gamma (x)\,\,, \eqno(2.19c) $$
where $ \mu_o $ is an arbitrary finite mass scale, $ \psi_{_{\ell}}^{\,0}(x) $ is
the free left-moving Fermi field given by
$$ \psi^{\,0}_{_{\!\!\ell}}(x)\,=\,\Bigl (\,\frac{\mu_o}{2\pi}\,\Bigr
)^{\,1/2}\,:\,e^{\,2i\sqrt \pi\,\phi_{_{\ell}}^{\prime}(x)}\,:\,\,\,, \eqno(2.20)
 $$
and 
$$ \Gamma (x)\,=\,:\,e^{\,2i\sqrt \pi\,(\,\frac{1}{\sqrt{a-1}}\,\chi_{_{\,2}}(x)\,
-\,\frac{g}{\sqrt \pi}\,\lambda^\prime(x)\,)}\,:\,\,.\eqno(2.21) $$
The Wick exponential $ : \exp \,i \Phi : $ has to be understood as a
formal series of Wick-ordered powers of the field $ \Phi $ at the
exponent\footnote{\it The problem for defining the Wick exponential of 
massless scalar fields in two
dimensions is that the corresponding Wightman functions do not
satisfy positivity unless some selection rule is introduced [19]. In
Ref.[17] the definition of the Wick exponential of the massless
scalar field in two dimensions as an operator-valued distribution is
discussed in the Krein space realization of the field.} [17]. 

The so-obtained fermionic fields (2.19b-c) satisfy abnormal
commutation relations. The Fermi field operators with correct
anticommutation relations can be obtained from the original set
by a Klein transformation [16]. In what follows we shall
suppress the Klein factors, since they are not needed for our present
purposes.

\subsection{Intrinsic Local Field Algebra}

The procedure which we shall adopt to display the basic structural properties 
of the model and obtain a consistent prescription to identify correctly its 
physical content, is to embed the mathematical structures of the bosonization 
scheme into the context of the general principles of Wightman field
theory. In this way the states that are admissible on the Hilbert
space $ {\cal H} $ are
selected according to some appropriate criteria and the physical 
interpretation of the theory is based on the
Wightman functions representing the intrinsic local field algebra 
which describes the degrees of freedom of the model. 

Within a precise mathematical point of view, a relativistic Quantum
Field Theory is formulated in terms of fields as local operator-valued
tempered distributions. Due to the singular character of the fields at a
point, only space-time averages of fields have physical meaning, in
general. In this way, in a Quantum Field Theory the role of the algebra of 
the canonical variables is played  by the {\it field algebra}
generated by polynomials of the smeared fields $ \Phi_i (f) $,
$$ \Phi_i (f)\,=\,\int\,\Phi_i (x)\,f(x)\,d^n x $$
with $ f(x) $ a smooth and fast decreasing regular test function.

A {\it local field algebra} $ \Im $ satisfies the algebraic
constraint given by locality,
$$ \bigl [\,\Phi (f)\,,\,\Phi (h)\,\bigr ]\,=\,0 $$
if supp $ f $ is space-like relative to supp $ h $. This property
expresses the fact that fields averages belonging to space-like
separated regions are simultaneously diagonalizable operators. Therefore, a
{\it local Quantum Field Theory} is specified by a representation of
the algebra $ \Im $ generated by the polynomials of the smoothed
local fields $ \Phi (f) $ satisfying the locality condition.

Given a QFT with a set of local fields $ \{ \Phi \} $. The set of
fields  $ \{ \Phi \} $ provides a complete description of the system
if for every space-like slab $ {\cal S} $ \footnote{\it A space-like
slab in space-time is defined as the open set between two parallel space-like
planes.}, the set of vectors {\Large $\wp$} $ ( \Phi (f) )\Psi_o $ spans the
Hilbert space $ {\cal H} $. Here {\Large $\wp$} is an arbitrary
polynomial in the $ \Phi (f) $, $ f $ having supports in $ {\cal S}
$. Equivalently, the set of fields $ \{ \Phi \} $ provides a complete
description of the system if for every space-like slab $ {\cal S} $, the 
sets  $ \{ \Phi (f) \} $, $ f $ with supp. in $ {\cal S} $, are
irreducible \footnote{\it The set of fields $ \{ \Phi (f) \} $ is
irreducible if every bounded operator which has the properties,
$$ {\cal O}\,{\cal D}\,\subset\,{\cal D}\,\,,$$
where $ {\cal D} $ is the common dense domain of the set $ \{ \Phi (f) \} $, and 
$$ \Phi (f)\,{\cal O}\,=\,{\cal O}\,\Phi (f)\,\,, $$
is a constant multiple of the identity.}.

The operator solution of the equations of motion defining the $GNI$
formulation of the CSM is given in terms of the {\it irreducible set of
field operators} $ \{\bar \psi, \psi, {\cal A}_{\mu}\} $ [11,16]. These 
field operators constitute the intrinsic mathematical description of
the model and serve as a kind of building material in terms of which the $GNI$
version of the model is formulated 
and whose Wightman functions define the model. Every operator of the
theory is a
function of the intrinsic set of field operators $ \{\bar \psi, \psi,
{\cal A}_{\mu}\} $ which defines a polynomial algebra $ \Im_{_{_{GNI}}} $, that
is, {\it the local field algebra $ \Im_{_{_{GNI}}} $ intrinsic to the $GNI$ 
formulation is generated  from the irreducible set of field operators
$ \{\bar \psi, \psi, {\cal A}_{\mu}\} $}, through polynomials of these 
smeared fields, Wick ordering, point-splitting regularization
of polynomials, etc. The Wightman functions 
generated from the intrinsic field algebra $ \Im_{_{_{GNI}}} $ identifies a vector 
space $ {\cal D}_o\,\equiv \,\Im_{_{_{GNI}}} \,\Psi _o $ of local states: the field algebra $ \Im_{_{_{GNI}}} $ is
represented in the Hilbert space $ {\cal H}_{_{_{GNI}}} \doteq \Im_{_{_{GNI}}} \,\Psi_o $ of
the $GNI$ formulation.

Within the bosonization scheme, the operator solution is constructed
in terms of Wick exponentials and derivatives of scalar fields in two 
dimensions. The introduction of the set of Bose 
fields $ \{\phi^{\prime}, \chi_{_2}, \lambda^{\prime}, \chi_{_1} \} $ defines 
an enlarged redundant field algebra $ \Im_{_{_{GNI}}}^{^B} $ which is represented on the
Hilbert space $ {\cal H}_{_{_{GNI}}}^{^B} \doteq \Im_{_{_{GNI}}}^{^B}\,\Psi_o $, which
provides a Fock-Krein representation\footnote{\it A careful and
rigorous treatment of the massless scalar field in two dimensions
requires the construction of a Krein space in which 
the set of 
fields $ \{\varphi\}\,\equiv\,\{\phi^{\prime}, \chi_{_2}, \lambda^{\prime}\} $ is
realized as a local operator-valued tempered distribution with dense
domain $ {\cal D}_o\,\equiv ${\large$\wp$}$(\varphi) \Psi_o $, where 
{\large$\wp$}$(\varphi)$ is the polynomial algebra generated by the
smoothed set of fields $ \{\varphi (f)\} $ [17]. In order to associate a 
Hilbert space of states to the Wightman functions we can follow along the same
lines as those of Ref.[11], and find a Krein topology which majorizes the
Wightman functions of $ \Im_{_{_{GNI}}} $ and defines a maximal Hilbert space 
structure.} of the algebra $ \Im_{_{_{GNI}}} ^{^B} $ [11]. The Bose fields  $ \{\phi^{\prime}, \chi_{_2}, \lambda^{\prime}, \chi_{_1} \} $
are the {\it building blocks} in terms of which the operator solution
is constructed and should not be considered as elements of the field algebra intrinsic to 
the model. Only some combinations of them belong to the intrinsic 
algebra $ \Im_{_{_{GNI}}} $. 

The field algebra $ \Im_{_{_{GNI}}} $ is a proper 
subalgebra of $ \Im_{_{_{GNI}}}^{^B} $ and
is recovered from a particular set of operators constructed from
linear combinations and Wick ordered exponentials of linear
combinations of Bose fields. Not all Bose fields
belong to the algebra $ \Im_{_{_{GNI}}} $ of the local fields, nor all vectors 
of $ {\cal H}_{_{_{GNI}}}^{^B} $ belong to the state space $ {\cal H}_{_{_{GNI}}} $ of the
$GNI$ formulation of the model. In this way, the set of local states $
{\cal D}_o^{^B} \,\equiv \, \Im_{_{_{GNI}}} ^{^B}\,\Psi _o $, corresponding to the largely
redundant field algebra $ \Im_{_{_{GNI}}} ^{^B} $, contains elements which are not
intrinsic to the model. Of course, the Hilbert space $ {\cal H}_{_{_{GNI}}} $ is a 
proper subspace of $ {\cal H}_{_{_{GNI}}}^{^B} $. 

Due to the presence of the longitudinal 
current $ L^{\,\mu} $ in Eq.(2.12), the Gauss' law holds in a weak
form and is satisfied on the physical 
subspace $ {\cal H}_{_{_{GNI}}}^{^{phys}} $ defined by the subsidiary 
condition $ L^{\,\nu}\,\approx\,0$, that is
$$\langle \,\Phi\,,\,(\,\,J^{\,\nu}(x)\,-\,\partial _{\mu}\,
{\cal F}^{\,\mu \nu}(x)\,)\,\Psi \,\rangle \,=\,\langle \,\Phi\,,\,L^{\,\nu }(x)\,\Psi\,\rangle\,=\,0\,\,\,\,\,,
\,\,\,\,\,\forall\, \Phi \,,\, \Psi \,\,\in\,\,
{\cal H}_{_{_{GNI}}}^{^{phys}}\,\,. \eqno (2.22)
$$
The algebra of the physical operators $ \Im_{_{_{GNI}}}^{^{phys}} $ must be
identified as the subalgebra of $ \Im_{_{_{GNI}}} $ which obeys the
subsidiary condition in a proper Hilbert space completion of the
local states. However, {\it it is a peculiarity of the
anomalous chiral model that the algebra $ \Im_{_{_{GNI}}}^{^{phys}} \equiv \Im_{_{_{GNI}}} $, since 
all operators belonging to the intrinsic local field algebra $ \Im_{_{_{GNI}}} $ commute 
with the longitudinal
current},
$$ [\,{\cal O}\,,\,L_{\mu }\,]\,=\,0\,\,\,,\,\forall\,{\cal O}\,\in
\,\Im_{_{_{GNI}}} \,\,\,.\eqno (2.23) $$
{\it This is expected in the anomalous case since the theory has lost the 
local gauge invariance}. Then the 
local field algebra $ \Im_{_{_{GNI}}} $ is a singlet under local gauge
transformations and is represented in the state space $ {\cal H}_{_{_{GNI}}}^{^{phys}}
\equiv {\cal H}_{_{_{GNI}}} $ of the $GNI$ formulation \footnote{ {\it This does not 
occur in the VSM, in which
neither $ \psi $ nor $ {\cal A}_\mu $  are singlet under local gauge 
transformations since they do not commute with the constraints. In this genuine 
gauge theory 
we have a prehilbert space so that by completions and quotients one gets a
physical Hilbert subspace defined by gauge invariant states accommodated 
as equivalent classes} [13].}. 

The intrinsic irreducible set of fields $ \{ \bar \psi, \psi, {\cal A}_\mu \} $, in terms 
of which the $GNI$ formulation is defined has been written in terms of the
building blocks $ \{ \chi_{_1}, \chi_{_2}, \phi ^\prime, \lambda ^\prime \} $ 
by Eqs.(2.19). From this set of intrinsic fields we construct all operators
belonging to the algebra $ \Im_{_{_{GNI}}} $, as for example the vector current
$$ J^{\,\mu}\,=\,\frac{1}{m_a}\,\tilde \partial ^{\,\mu}\chi_{_1}\,+\,L^{\,\mu}
\,\,,\eqno(2.24) $$
and the field-strength tensor
$$ {\cal F}^{\,\mu\nu}\,=\,m_a\,\varepsilon^{\,\mu\nu}\chi_{_1}\,\,.
\eqno(2.25) $$
Since the field $ \chi_{_1} $ belongs to $ \Im_{_{_{GNI}}} $, i.e.,
$$ \chi_{_1}\,=\,\frac{1}{2m_a}\,\varepsilon^{\,\mu\nu}{\cal F}_{\mu\nu}\,\,\,
\in\,\Im_{_{_{GNI}}} \,\,,\eqno(2.26) $$
then, besides the longitudinal current, the Wick exponential $ :\exp\,\{2i\sqrt{\pi/(a-1)}\,\chi_{_1}\,\}:\,
\in \,\Im_{_{_{GNI}}} $ [11,17]. Therefore the operator
$$ W\,\doteq\,:\,\exp \,\{\,2i\sqrt{\frac{\pi}{a-1}}\,
\chi_{_{1}}\,\}\,\psi_{_{\!\!\ell}}\,:\,=\,
\psi^{\,0}_{_{\!\!\ell}}\,\Gamma\,\,,\eqno(2.27) $$
also belongs to $ \Im_{_{_{GNI}}} $. {\it In order to be defined on $ {\cal H}_{_{_{GNI}}} $ the operator $ W $ cannot be
further reduced}. $ \Gamma $ is a spurious operator with zero scale dimension
and spin that generates constant Wightman functions
$$ \langle \,\Gamma^*(x_1) \cdots 
\Gamma^*(x_m) \Gamma^*(x_{m+1})\cdots \Gamma^*(x_n)\,\Gamma  (y_1) \cdots 
\Gamma (y_m) \Gamma (y_{m+1})\cdots \Gamma (y_n)\,\rangle = 1 \,\,,
\forall \,a > 1\,\,, \eqno (2.28)$$
and therefore the correlation functions of the operator $ W $ are
isomorphic to those of the free fermion field $ \psi^{\,0}_{_{\!\!\ell}} $,
$$ \langle\,\prod _{i=1}^n\,W^*(x_i)\,\prod_{j=1}^n\,W(y_j)\,\rangle\,\equiv\,
 \langle\,\prod _{i=1}^n\,\psi^{*\,0}_{_{\!\!\ell}}(x_i)\,\prod _{j=1}^n\,
 \psi^{\,0}_{_{\!\!\ell}}(y_j)\,\rangle\,\,\,,\,\,\,\forall\,\,a > 1\,\,.\eqno (2.29) $$

The Hilbert space $ {\cal H}_{_{_{GNI}}} $ of the $GNI$ formulation can be factorized as a product
$$ {\cal H}_{_{_{GNI}}} =  {\cal H}_{\chi_1} \otimes {\cal H}_{\psi^o,\,
{\chi_2},\,\lambda^{\prime}}\,\,, \eqno (2.30)$$
where $  {\cal H}_{\chi_1} $  is the Fock space of the free massive
field $ \chi_1\,\in \,\Im_{_{_{GNI}}} $ and $ {\cal H}_{\psi^o,\,
{\chi_2},\,\lambda^{\prime}} $ the closure of the space
$$  {\cal H}_{\psi^o,\,{\chi_2},\,\lambda^{\prime}} = \Im_{_{_{GNI}}} ^{\prime}\,\Psi_o \,\,,\eqno (2.31)$$
in which we have a representation of the field algebra $ \Im_{_{_{GNI}}}
^{\prime} $ generated by $ j^{\mu}_{_{\ell}}$, $ L ^\mu $,  $ W $ and
the longitudinal piece of the gauge field $ {\cal A}^\mu_{_{L}} $
(which only depends on the massless fields); i. e., the closure of the space is obtained by applying to 
the vacuum polynomials of the 
fields $ j^\mu_{_{\ell}} $, $ W $, $ L ^\mu $ and $ {\cal
A}^\mu_{_{L}} $. Except for the Wick exponential $ W $, the field
algebra $ \Im^\prime_{_{_{GNI}}} $ is generated by  combinations of 
derivatives of
the fields $ \phi ^\prime, \chi _2 $ and $ \lambda ^\prime $. The
Hilbert space $ {\cal H}_{_{_{GNI}}} $ is given by
$$ {\cal H}_{_{_{GNI}}}\,\equiv \,\Im (\,\chi_{_1},\Im^\prime_o\,
)_{_{_{GNI}}}\Psi_o\,\,,\eqno (2.32) $$
where $ \Im (\,\chi_{_1},\Im^\prime_o\,)_{_{_{GNI}}} $ is the field algebra
generated by $ \chi_{_1} $ and by the subalgebra $ \Im^\prime_o\,\subset\,
\Im_{_{_{GNI}}}^\prime $ which commutes with $ L^\mu $.

The operator $ W $ cannot be 
reduced and the Hilbert space 
completion $ {\cal H}_{\psi^o,\,{\chi_2},\,\lambda^{\prime}} $ cannot be 
further decomposed. The Hilbert space factorization (2.30) and the
isomorphism expressed by Eq.(2.29) imply that the Hilbert space $
{\cal H}_{_{_{GNI}}} $ contains free fermion states. As we shall see 
later, the improper factorization of the 
completion of states performed in 
Refs.[9,10] leads to some misleading conclusions about basic structural properties of the model.

\subsection{Cluster Decomposition}

Although the field operator $ \psi_{_{\!\!\ell}} $ representing the fermionic
particle in the $GNI$ formulation is given in terms of a spurious
operator $ \Gamma $, the cluster property for the corresponding
Wightman functions is not violated. In particular, we obtain for the
fermion two-point function [8,12]
$$ \langle\,\psi^*_{_{\!\!\ell}}(x)\,\psi_{_{\!\!\ell}}(y)\,\rangle \,=\,\,\langle\,
\psi^{*\,0}_{_{\!\!\ell}}(x)\,\psi^{\,0}_{_{\!\!\ell}}(y)\,\rangle \,\exp \,\{\,\frac{4\pi}{a-1}\,\Delta^{(+)}(x-y; m_a)\,\} \,\,.
\eqno(2.33) $$

Contrary to what happens in the VSM [13], the space-time contribution
coming from the free fermion two-point function in eqs.(2.29-33) ensures
the existence of fermions in the asymptotic states and implies that
the cluster decomposition is not violated
$$  \langle \,\psi_{_{\!\!\ell}} (x + \lambda)\Psi_o\,,\,
\psi _{_{\!\!\ell}} (x) \Psi_o\,
 \rangle  _{\phantom{}_{\phantom{}_{ {\lambda \rightarrow \infty }}}}
\,\longrightarrow \, 0\, =\, \vert\!\vert \,\langle\,\Psi_o\,,\,\psi
_{_{\!\!\ell}}(x)\,\Psi_o \,\rangle\,\vert\!\vert ^{\,2}
\,\,. \eqno (2.34)$$

Of course, we can construct a composite operator belonging to $
\Im_{_{_{GNI}}} $ that carries the free fermion chirality,
$$ M\,=\,\psi_{_{\!\!\ell}}^*\,\psi_{_{\!\!r}}\,=\,\frac{\mu_o}{4\pi}\,:\,e^{\,2i\sqrt\pi\,
\tilde\phi}\,:\,=\,:\,e^{\,2i\sqrt{\frac{\pi}{a-1}}\,\chi_{_1}}\,:\psi^{\,0*}_{_{\!\!\ell}}\,
\psi^{\,0}_{_{\!\!r}}\,\Gamma^* \,\,\,\in \,\Im_{_{_{GNI}}} \,\,.\eqno(2.35) $$
However the cluster decomposition is not violated for the
corresponding Wightman functions as well :
$$ \langle \,M^{\ast}(x)\,M(0)\,\rangle\,=\, e^{\,\frac{4 \pi}{a - 1} \Delta ^{(+)}(x; m_a)}\,
  \Biggl ( \frac{\mu _o}{2 \pi x^{\,2}} \Biggr )_{x^2 \rightarrow - \infty} 
  \rightarrow  0 \,\,. \eqno (2.36)$$ 
From the cluster property (2.34) and (2.36) we conclude
that {\it the screening and confinement aspects exhibited by the CSM 
for $ a > 1 $ differ from those of the VSM} [13]. This result is in agreement 
with 
the conclusions of Refs.[8,12], and is crucial for the discussion
concerning the claimed equivalence of the VSM and CSM defined for $ a = 2 $.

A peculiar feature of the CSM which differs from the vector case
is the fact that {\it the cluster decomposition property is not violated
for Wightman functions that are representations of the field 
algebra $ \Im_{_{_{GNI}}}  $ generated from the irreducible set of field 
operators $ \{\,\bar\psi, \psi, {\cal A}_{\mu}\,\} $, provided
only the intrinsic field algebra is considered without any reduction
of the completion of states}. Although the cluster
property is violated for the Wightman functions of the 
operator $ \Gamma $, this operator cannot be defined 
in $ {\cal H}_{_{_{GNI}}} $ since $ [\,{\cal Q}^{^5}_{_L}\,,\,\Gamma (x)\,]\,\neq \,0
$, where $ {\cal Q}^{^5}_{_L} $ is the chiral charge associated with the
longitudinal current.

\subsection{Connection with the Minimal Bosonization Scheme}

Now, let us show that the minimal bosonization scheme used in 
Ref.[12] is physically equivalent to the enlarged bosonization scheme
used in Refs.[8,9], in spite of the fact that the latter introduces an 
additional redundant field algebra.

The field operator $ \psi_{_{\!\!\ell}} $ given by eq.(2.19c) can be related to the
operator solution $ \hat \psi_{_{\!\!\ell}} $ given in Ref.[12] in the
following way: Defining the non-canonical scalar field
$$ \sigma (x) \,\doteq\,\frac{1}{\sqrt{a - 1}}\,\chi_{_1}(x)\,\,, \eqno(2.37) $$
we can write $  \psi_{_{\!\!\ell}}(x) $ as
$$ \psi_{_{\!\!\ell}}(x)\,=\,\hat\psi_{_{\!\!\ell}}(x)\,
\Gamma (x) \,\,,\eqno(2.38) $$
where the fermion field operator in the minimal bosonization scheme
is given by \footnote{ {\it The canonical harmonic field $ h (x) $ used in 
Refs.[1,12] is identified with [8]}

$$ h_{_{\ell}}(x)\,\equiv \, \phi^{\prime}_{_{\ell}}(x)\,+\,\Bigl (\,
\frac{1}{\sqrt {a - 1}}\chi_{_2}(x)\,-\,\frac{g}{\sqrt \pi}\,\lambda
^{\prime}(x)\,\Bigr ) $$

\noindent {\it such that, and in accordance with the results obtained in 
Refs.[1,2,5,6,8,12], there are only two ``physical'' bosonic degrees of freedom in 
the anomalous model: the massive field $ \chi_{_{1}} $ and the massless 
field $ \phi ^{\prime} $, which acts as potential for the free fermionic 
current.}}
$$ \hat\psi_{_{\!\!\ell}}(x)\,= :e^{- 2 i \sqrt \pi \sigma (x)} :\,
\psi_{_{\!\!\ell}}^o (x)\,\,.  \eqno(2.39) $$
The fact that the $GNI$ operator solutions $ \hat \psi (x) $ and $ \psi (x) $ 
are related by a superfluous spurious phase operator means that 
the bosonization scheme used in Refs.[8,9] introduces an 
extra (unphysical) redundant field algebra, i. e., it introduces more 
degrees of freedom than those needed for the description of the model. Of 
course, {\it these operator solutions are  equivalent since they generate 
the same Wightman functions, provided only the 
intrinsic field algebra is considered without any reduction of the
completion of states}.

The Hilbert space $ \hat {\cal H}_{_{_{GNI}}} $ of the minimal scheme
can be decomposed as
$$  \hat {\cal H}_{_{_{GNI}}} = {\cal H}_{\sigma} \otimes {\cal H}_{\psi
^o}\,\,,\eqno (2.40) $$
and the isomorphism expressed in Eq.(2.29) means that {\it the Wightman
functions that are represented on the Hilbert space 
completion $ {\cal H}_{\psi^o} $ of the minimal scheme are
isomorphic to those which are represented on the Hilbert space 
completion
$ {\cal H}_{{\chi_2},\,\lambda^{\prime},\,\psi_o} $ of the local states of the 
enlarged $GNI$ formulation}. This ensures the existence of fermions in
the asymptotic states \footnote{ \it In 
the VSM the field $ \lambda^\prime $ is 
absent, $ \chi_{_1}\,\equiv \,\tilde \Sigma $ and $ \chi_{_2}\,\equiv\,
\tilde \eta $. The Hilbert space can be decomposed as [11]

$$  {\cal H}_{_{_{Sch}}} = {\cal H}_{\tilde \Sigma} \otimes {\cal H}_{\psi^o,\,
\tilde \eta}\,\,. $$}.

\subsection{The case $ a = 2 $}

This particular case has generated some confusion in the literature, since 
the corresponding operator algebra exhibits non-trivial and delicated features 
which might lead to misleading conclusions about structural  properties of the 
model. As we shall see, one can construct a field 
subalgebra $ \Im_{_{Sch}}\,\subset \,^{^B}\Im_{_{_{GNI}}}^{^{ext}} $ which is 
isomorphic to the field subalgebra of
the VSM but does not belong to the intrinsic field algebra $ \Im_{_{_{GNI}}} $ of 
the anomalous chiral model \footnote{ The introduction of the dual fields $ \tilde \lambda
^\prime $ (or $ \tilde \chi _{_2} $) enlarges the bose field 
algebra $ \Im_{_{_{GNI}}} ^{^B} $ and
defines an external algebra  $ ^{^B}\Im_{_{_{GNI}}}^{^{ext}}\,\supset \,\Im_{_{_{GNI}}}
^{^B} $.}.

For $ a = 2 $ the Fermi field operator $ \psi_{_{\!\!\ell}} $ can be 
factorized as
$$  \psi _{_{\!\!\ell}} =\,\Bigl (\,\frac{\mu_o}{2\pi}\,\Bigr )^{1/2}\,
:e^{\,- 2 i \sqrt \pi \chi _{_1}}:\,\sigma _{_{\!\!\ell}} \,\sigma ^{\,\ast}_{_{\!\!r}} 
:e^{\,- 2 i g \lambda^{\,\prime}_{_{\ell}}}:\,\,\in \,\Im_{_{_{GNI}}} \,\,, \eqno (2.41)$$
where $ \sigma _{_{\!\!r}} $ and $ \sigma _{_{\!\!\ell}} $ are spurious 
operators, and as we shall prove later they do not belong to $ \Im_{_{_{GNI}}} $, given 
by 
$$ \sigma _{_{\!\!r}} = :e^{\,-\,i \sqrt \pi \,[\,(\chi_{_2}\,+\,
\tilde \chi_{_2}\,)\,-\,\frac{g}{\sqrt
\pi}\,(\,\lambda^{\,\prime}\,+\,\tilde \lambda^{\,\prime}\,)\,]}:\,
 =\, :e^{\,i \sqrt \pi L_{_{r}}}:\,\,\,,  \eqno (2.42a) $$ 
$$ \sigma _{_{\!\!\ell}} =  :e^{\,i \sqrt \pi \,[\,
(\phi ^{\,\prime}\,-\,\tilde \phi ^{\,\prime})\,+\,(\,\chi
_{_2}\,-\,\tilde \chi _{_2}\,)\,]}: = :e^{\,i \sqrt \pi L_{_{\ell}}}:\,\,\,. 
 \eqno (2.42b)$$ 

For $ a = 2 $, the field $ \lambda^{\prime}_{_{\!\ell}} $ decouples 
from the longitudinal current $ L_{_{\!\ell}} $ given by Eq.(2.17d)
and there is a broader class of operators belonging to $
\Im^{^B}_{_{_{GNI}}} $ that satisfy the subsidiary condition. Since in this 
case the 
Wick exponential operator $ :\exp \{- 2 i g \lambda^\prime_{_{\ell}}\}: $ commutes with 
the constraints \footnote{ \it Where we used $ \lambda_{_{\ell}}\,=\,1/2\,(\,
\lambda^\prime\,-\,\tilde \lambda^\prime\,) $.}
$$ [\,L_\mu \,,\,:e^{\,i\,g\,(\tilde \lambda^\prime
\,-\,\lambda^\prime)}:\,]\,=\,0 \,\,, \eqno (2.43)$$
it is tempting to extract from the operator (2.41) the dependence on the
field $ \lambda^\prime_{_{\ell}} $, as is done in Ref.[9] for the corresponding
$GI$ field operator. The
resulting operator shares some resemblance with the composite chiral
operator of the VSM [9]. With this procedure the authors of Ref.[9] conclude
for the need of the $ \Theta $-vacuum parametrization in the
anomalous model and its equivalence with the VSM. As we shall see, in the formulation in terms of intrinsic 
field algebra
generated from the irreducible set of local field operators, the
operators 
$ \sigma _{_{\!\!\ell}} \,\sigma ^{\,\ast}_{_{\!\!r}} $, $ \sigma_{_{\ell}} $, 
$ \sigma _{_r} $ and $:e^{\,i\,g\,(\tilde \lambda^\prime
\,\mp\,\lambda^\prime)}: $  do not exist separately and
cannot be defined on $ {\cal H}_{_{_{GNI}}} $.

In order to show that some misleading conclusion can arise by making
the factorization of the closure space, as for 
example $ {\cal H}_{\psi^o,\,{\chi_2},\,\lambda^{\prime}}\,=\,
{\cal H}_{\psi^o,\,{\chi_2},\,\lambda^\prime_{_r}}
\otimes {\cal H}_{\lambda^{\prime}_{_{\ell}}} $ done in Ref.[9], we follow
along the same lines as those of Ref.[9] and consider the 
operator $ \psi^{\prime}_{_{\!\!\ell}}(x)\,\in \,\Im_{_{_{GNI}}} ^{^B} $ defined for
$ a > 1 $ and given by 
$$ \psi^{\prime}_{_{\!\!\ell}}(x)\,\doteq :\,e^{+\,i\,g\,[\,\lambda
^\prime(x)\,-\,\tilde \lambda^\prime(x)\,]}\,
 \psi_{_{\!\!\ell}}(x)\,:\,\notin \,\Im_{_{_{GNI}}} \,\,,\eqno (2.44)$$
Then, we construct the composite operator carrying free chirality
$$ S\,=\,{\psi^{\prime}_{_{\!\!\ell}}}^*\,\psi_{_{\!\!r}}\,\,\,.\eqno (2.45)$$
The operator $ S\,\in \,\Im_{_{_{GNI}}}^{^B} $ is the $GNI$ counterpart of the operator 
which was considered in Ref.[9] in the case $ a = 2 $ and the corresponding 
two-point function is given by
$$ \langle \,S^{\,\ast}(x)\,S (0)\,\rangle\,=\, e^{\,\frac{4 \pi}{a - 1} \Delta ^{(+)}(x;\, m_a)}\,
 \frac{\mu_0}{2 \pi}\Bigl ( x^{\,2} \Bigr )^{-\,(a - 2)/(a -
1)}\,\,.\eqno (2.46)$$
{\it For $ a = 2 $ the cluster decomposition is violated !} Of course,
this implies a misleading conclusion about the need of the $ \Theta $-vacuum 
parametrization in the $GNI$ formulation of the theory. {\it The state $
\psi^{\prime}_{_{\!\!\ell}}\,\Psi_o $ belongs to the
improper Hilbert space
completion $ {\cal H}_{\psi^o,\,{\chi_2},\,\lambda^\prime_{_r}}\,\subset \,
{\cal H}_{_{_{GNI}}}^{^B} $ of the local 
states and cannot be regarded as a state in
the Hilbert space of states which defines the representation of the
field algebra $ \Im_{_{_{GNI}}} $}. As we shall see, {\it the set of local 
states $ {\cal D}_o\,\equiv \,\Im_{_{_{GNI}}} \,\Psi_o $ does not contain states 
like $ \psi^{\prime}_{_{\!\!\ell}}\,\Psi_o $}. 

Note that the introduction of the dual field $ \tilde \lambda
^\prime $ (or $ \tilde \chi _{_2} $) enlarges the bose field 
algebra $ \Im_{_{_{GNI}}} ^{^B} $ and
defines an external algebra  $ ^{^B}\Im_{_{_{GNI}}}^{^{ext}}\,
\supset \,\Im_{_{_{GNI}}} ^{^B} $. The need 
of the $ \Theta $-vacuum parametrization, claimed in Ref.[9], cannot be regarded as a structural 
property of the theory since it emerges as a consequence  of the use of the redundant
algebra belonging to the external algebra $ ^{^B}\Im_{_{_{GNI}}}^{^{ext}} $, generated
by the set of Bose fields $ \{\phi^{\prime}, \tilde \phi^{\prime}, \chi_{_2}, 
\tilde \chi_{_2}, \lambda^{\prime}, \tilde \lambda^{\prime},
\chi_{_1} \} $, rather than on the 
intrinsic algebra $ \Im_{_{_{GNI}}} $, generated from
the irreducible set of field operators which defines the model. 

We must remark that on the enlarged state space $ {\cal H}_{_{_{GNI}}}^{^B} $, the connection with 
the VSM can be made by considering the operator $ S_{_{Sch}}\,\doteq\,
\psi^\prime_{_{\!\!\ell}}\,\vert_{_{a=2}} $
$$ S_{_{Sch}}\,\doteq\,:\,e^{\,+\,2\,i\,g\,\lambda_{_{\ell}}^{\prime}}\,
\psi_{_{\!\!\ell}}\,:\,\vert_{_{a=2}}\,=\,
\Bigl (\,\frac{\mu_o}{2\pi}\,\Bigr )^{1/2}\,
:e^{\,- 2 i \sqrt \pi \chi _{_1}}:\,\sigma _{_{\!\!\ell}} \,\sigma ^{\,\ast}_{_{\!\!r}} 
\,\,\not\in \,\,\Im_{_{_{GNI}}} \,\,,\eqno (2.47)$$
such that the cluster property is violated for the corresponding
Wightman function
$$ \langle\,S^*_{_{Sch}}(x)\,S_{_{Sch}}(0)\,\rangle\,_{_{x^2\,
\rightarrow\,-\,\infty}}\,\rightarrow\,\frac{\mu_o}{2\pi}\,\,. \eqno (2.48) $$

The operator $ S_{_{Sch}}\,\not\in\,\Im_{_{_{GNI}}} $ and can be viewed as a composite operator 
$$ S_{_{Sch}}(x)\,=\,\lim_{\epsilon\,\rightarrow\,0}\,{\cal Z}^{\,- 1}(\epsilon)\,
\Psi^{\,\ast}_{_{\!\!r}}(x\,+\,\epsilon)\,\Psi_{_{\!\!\ell}}(x)\,\,, \eqno(2.49) $$
where $ \Psi \,\not\in\,\Im_{_{_{GNI}}} $ and is given by
$$ \Psi (x)\,=\,\Bigl (\,\frac{\mu_o}{2\pi}\,\Bigr )^{\,1/4}\,:\,e^{\,i\,
\sqrt\pi\,\gamma^{\,5}\,\chi_{_1}(x)}\,:\,e^{\,i\,\sqrt\pi\,[\,\gamma^{\,5}\,
\tilde L(x)\,+\,L(x)\,]}\,:\,\,\,. \eqno(2.50) $$

Although the operator $ \Psi (x) $ generates the same Wightman functions
of the covariant solution of the VSM, it is not the
intrinsic field operator representing the fermionic content of the
CSM and cannot be defined on $ {\cal H}_{_{_{GNI}}} $. The operators (2.47) and
(2.50) are the gauge noninvariant counterpart of the field operators used
in Refs.[9,10] to suggest the need of $ \Theta $-vacuum
parametrization and the equivalence of the VSM and CSM defined for $ a
= 2 $. Since these operators do not belong to the field algebra $ \Im_{_{_{GNI}}}
$, they cannot be defined on $ {\cal H}_{_{_{GNI}}} $ and the claimed equivalence
with the VSM cannot be established. This apparent equivalence is a
by-product of the incorrect choice of the observables representing
the fermionic content of the model, that arises by factorizing
the closure of the space as the 
product $ {\cal H}_{\psi^o,\,{\chi_2},\,\lambda^{\prime}}\,=\,
{\cal H}_{\psi^o,\,{\chi_2},\,\lambda^\prime_{_r}} 
\otimes {\cal H}_{\lambda^{\prime}_{_{\ell}}} $. Note that, in
the VSM [11], $ \chi_2 \equiv \tilde \eta $\,,\,$ \chi_1 \equiv \tilde \Sigma $
and the closure of the space is  $ {\cal H}_{\psi^o,\,\tilde \eta} $
and cannot be factorized as   $ {\cal H}_{\psi^o,\,\tilde \eta} = 
{\cal H}_{\tilde \eta} \otimes {\cal H}_{\psi_o} $. The Hilbert space
of the VSM can be identified with the improper 
subspace $ {\cal H}_{Sch} \equiv {\cal H}_{\chi_1} \otimes 
{\cal H}_{\psi^o_{_{\!\!\ell}},\,{\chi_2},\,\lambda^\prime_{_r}} $ of the chiral 
model defined for $ a = 2 $. 

Now we prove the proposition according to which one cannot define on
the Hilbert space $ {\cal H}_{_{_{GNI}}} $ the Wick exponential 
$$ \Lambda_{_{\ell}}(x)\,\doteq\,: e^{\,i\,g\,[\,\lambda^\prime(x)\,-\,
\tilde \lambda^\prime(x)\,]}:\,\,,\eqno (2.51) $$
and therefore the operator $ \psi^{\prime}_{_{\!\!\ell}}\,\doteq \,
:\Lambda_{_{\ell}}\,\psi_{_{\!\!\ell}}: $ also cannot be defined on $
{\cal H}_{_{_{GNI}}}$. In analogy
with the vector case [11], such property follows from the fact that
some charges get trivialized in the restriction 
from $ ^{^B}{\cal H}_{_{_{GNI}}}^{^{ext}} $
or from $ {\cal H}_{_{_{GNI}}}^{^B} $ to $ {\cal H}_{_{_{GNI}}} $.

For $ a \neq 2 $, the trivialization of the chiral 
charge $ {\cal Q}^{^5}_{_L} $ in the
restriction from $ {\cal H}_{_{_{GNI}}}^{^B} $ to $ {\cal H}_{_{_{GNI}}} $,
$$  {\cal Q}^{^5}_{_L}\,{\cal H}_{_{_{GNI}}}^{^B}\,\neq\,0\,\,\,,\,\,\,
{\cal Q}^{^5}_{_L}\,{\cal H}_{_{_{GNI}}}\,=\,0\,\,, \eqno (2.52) $$
implies that the closure of local states associated to the intrinsic field 
algebra $ \Im_{_{_{GNI}}} $ does not allow the introduction of operators which are charged 
under $ {\cal Q}^{^5}_{_L} $. However, for the especial case $ a = 2 $, the
field operators $ \Lambda_{_{\ell}}(x) $ and $ \psi_{_{\ell}}^\prime (x) $ are
neutral under $ {\cal Q}^{^5}_{_L} $,
$$ \bigl [\,{\cal Q}^{^5}_{_L}\,,\,\Lambda_{_{\ell}}(x)\,\bigr
]\,=\,0\,\,, \eqno (2.53)$$
$$ \bigl [\,{\cal Q}^{^5}_{_L}\,,\,\psi_{_{\ell}}^\prime (x)\,\bigr
]\,=\,0\,\,,\eqno (2.54) $$
and the criterion based on the trivialization of the 
charge  $ {\cal Q}^{^5}_{_L} $ is insufficient to decide about the existence
of the state $ \psi_{_{\ell}}^\prime\,\Psi_o $ on $ {\cal H}_{_{_{GNI}}} $.

Consider a local charge operator 
$$ {\cal Q}_{_{\tilde\varphi}}\,\doteq\,\lim _{R \rightarrow \infty}\,\int
_{_{_{\vert x^{ 1} \vert \leq R }}}\,\partial_{_{x^0}}\,\tilde\varphi
(x^0,x^1)\,d x^1\,\equiv \,\lim _{R \rightarrow \infty }\,
{\cal Q}_{_{\tilde\varphi,R}}\,\, \eqno (2.55) $$
associated with the field 
operator $ \tilde\varphi (x) $ \footnote{\it In which a regularization of the
integral is to be understood [11],
$$ {\cal Q}_{_{\tilde\varphi,R}}\,=\,\int\,\partial_{_{x^0}}\,\tilde\varphi
(x^0,x^1)\,f_{_R}(x^1)\,g(x^0)\,d x^1\,d x^0\,\,, $$
with $ f_{_R} = f(\vert x^1 \vert/ R) $, $ f \in {\cal D} (\Re^1) $, $ f(x) 
= 1 $ for $ \vert x \vert < 1 $, $ f (x) = 0 $ for $ \vert x \vert > 1 + 
\epsilon $, $ g (x^0) \in {\cal D} (\Re^1) $, $ \int g(x^0) dx^0 = 1 $.}
such that it get trivialized in the restriction 
from $ ^{^B}{\cal H}_{_{_{GNI}}}^{^{ext}} $ or 
from $ {\cal H}_{_{_{GNI}}}^{^B} $ to $ {\cal H}_{_{_{GNI}}} $, i.e.,
$$ {\cal Q}_{_{\tilde\varphi}}\,^{^B}{\cal H}_{_{_{GNI}}}^{^{ext}}\,\neq \,0
\,\,\,\,\,\,,\,\,\,\,\,\,
\cases { {\cal Q}_{_{\tilde\varphi}}\,{\cal H}_{_{_{GNI}}}^{^B}\,\neq\,0\cr or \cr
 {\cal Q}_{_{\tilde\varphi}}\,{\cal H}_{_{_{GNI}}}^{^B}\,=\,0 \cr}\,\,\,\,\,\,,\,\,
\,\,\,\,{\cal Q}_{_{\tilde\varphi}}\,{\cal H}_{_{_{GNI}}}\,=\,0 \,\,\,.
\eqno (2.56)$$

The trivialization of the charge $ {\cal Q}_{_{\tilde\varphi}} $ on
the Hilbert 
space $ {\cal H}\,\doteq\,\Im \,\Psi_o $,
follows from the fact that [11], if the field algebra $ \Im $ contains 
functions
of $ \varphi $ but not of $ \tilde\varphi $, then
$$ \langle\,{\cal Q}_{_{\tilde\varphi}}\Psi_o\,,\,\Im\,\Psi_o\,
\rangle\,=\,0\,\,, \eqno(2.57)$$
and hence, we get 
$$ \lim_{R \rightarrow \infty}\,\langle\,{\cal Q}_{_{\tilde \varphi,R}}\,
\Im\,\Psi_o\,,\,\Im\,\Psi_o\,\rangle\,= $$
$$ =\,\lim_{R \rightarrow \infty}\,\langle\,\bigl [\,
{\cal Q}_{_{\tilde \varphi,R}}\,,\,\Im\,\bigr ]\,\Psi_o\,,
\,\Im\,\Psi_o\,\rangle\,+\,
\lim_{R \rightarrow \infty}\,\langle\,\Im\,{\cal Q}_{_{\tilde \varphi,R}}\,
\Psi_o\,,\,\Im\,\Psi_o\,\rangle\,=\,
0 \,\,,\eqno (2.58)$$
i.e., for the set of local states $ {\cal D}_o\,
\equiv\,\Im_{_{_{GNI}}}\Psi_o $, we obtain the weak limit
$$ {\cal Q}_{_{\tilde\varphi}}{\cal D}_o\,=\,w - \lim_{R \rightarrow \infty}\,
{\cal Q}_{_{\tilde\varphi,R}}{\cal D}_o\,=\,0\,\,. \eqno (2.59) $$

Now, we prove that the trivialization of the 
charge $ {\cal Q}_{_{\tilde\lambda^\prime}} $ in the restriction 
from $ ^{^B}{\cal H}^{^{ext}}_{_{_{GNI}}} $ to $ {\cal H}_{_{_{GNI}}} $ implies
that the state $ \psi^{\prime}_{_{\!\!\ell}}(x)\,\Psi _o $ cannot belong 
to $ {\cal H}_{_{_{GNI}}} $. 

To this end, let $ A (\tilde\varphi ) $ be an element of the local field 
algebra $ \Im^{^B}_{_{_{GNI}}} $ (is a local operator) and that
have a non zero charge $ {\cal Q}_{_{\tilde\varphi}} $,
$$ \lim _{R \rightarrow \infty} \,
\bigl [\,{\cal Q}_{_{\tilde\varphi,R}}\,,\,A(\tilde\varphi
)\,\bigr ]\,=\,\alpha \,A(\tilde\varphi )\,\,\,,\,\,\,\alpha \,\neq \,0\,\,\,,
\,\,\, A(\tilde \varphi)\,\in\,\Im^{^B}_{_{_{GNI}}}\,\,.\eqno (2.60) $$
For the set of local 
states $ {\cal D}_o\,\equiv\,\Im_{_{_{GNI}}}\,\Psi_o $, and a local 
state $ A (\tilde\varphi ) \Psi_o $ of charge $ \alpha $, consider
$$ \alpha \,\langle\,A(\tilde\varphi ) \Psi_o\,,\,\Im_{_{_{GNI}}}\,
\Psi_o\,\rangle\,=\,\langle\,
\bigl [\,{\cal Q}_{_{\tilde\varphi}}\,,\,A(\tilde\varphi)\,\bigr]\,
\Psi_o\,,\,\Im_{_{_{GNI}}}\,\Psi_o\,
\rangle\,=$$
$$=\,\langle\,A(\tilde\varphi )\,\Psi_o\,,\,{\cal Q}_{_{\tilde\varphi}}\,
\Im_{_{_{GNI}}}\,\Psi_o\,\rangle\,-\,\langle\,
A(\tilde\varphi )\,{\cal Q}_{_{\tilde\varphi}}
\,\Psi_o\,,\,\Im_{_{_{GNI}}}\,\Psi_o\,\rangle \,=\,0\,\,,\eqno (2.61)
$$
and therefore the state $ A(\tilde\varphi )\,\Psi_o $ does not exist 
in $ {\cal H}_{_{_{GNI}}} $. 

Considering $ \tilde\varphi \equiv \tilde \lambda^\prime $, Eq.(2.61)
implies 
that {\it the closure of the local states associated to the 
field algebra $ \Im_{_{_{GNI}}} $ intrinsic to the model does not allow the 
introduction of operators which are charged under $ {\cal Q}_{_{\tilde \lambda
^\prime}} $}. Since
$$ \bigl [\,{\cal Q}_{_{\tilde \lambda^\prime}}\,,\,\Lambda_{_{\ell}}(x)\,
\bigr ]\,=\,\frac{\pi}{g}\,\Lambda_{_{\ell}}(x)\,\,, \eqno (2.62a) $$
$$ \bigl [\,{\cal Q}_{_{\tilde \lambda^\prime}}\,,\,
\psi^{\prime}_{_{\!\!\ell}}(x)\,\bigr ]\,=\,\frac{\pi}{g}\,
\psi^{\prime}_{_{\!\!\ell}}(x)\,\,, \eqno (2.62b) $$
{\it the set of local states $ {\cal D}_o \equiv \Im_{_{_{GNI}}} \,\Psi _o
$ does not contain states like $ \Lambda_{_{\ell}}(x)\,\Psi _o $ and
$ \psi^{\prime}_{_{\!\!\ell}}(x)\,\Psi _o $}. The state $ \psi _{_{\ell}}^\prime \,\Psi _o $
cannot belong to $ {\cal H}_{_{_{GNI}}} $ and the 
operator $ \psi _{_{\ell}} ^\prime \,\not\in \,\Im_{_{_{GNI}}} $.

Similarly, for $ \tilde\varphi\,\equiv\,\tilde
\chi_{_2}\,\in\,^{^B}\Im_{_{_{GNI}}}^{^{ext}} $ and the trivialization of the
charge $ {\cal Q}_{_{\tilde \chi_{_2}}} $ in the restriction 
from $ ^{^B}{\cal H}^{^{ext}}_{_{_{GNI}}} $ to $ {\cal H}_{_{_{GNI}}} $, one proves that the
operators $ \sigma_{_{\ell}} $ and $ \sigma _{_r} $ which are charged under $
{\cal Q}_{_{\tilde \chi_{_2}}} $, do not exist separately in $ \Im_{_{_{GNI}}} $ and
cannot be defined on $ {\cal H}_{_{_{GNI}}} $. Only the neutral combination $ W $
exist in the field algebra $ \Im_{_{_{GNI}}} $.

In conclusion we showed that the main results of Refs.[9,10] cannot be regarded as 
structural properties of the CSM since they are consequence of the use of a
redundant field algebra belonging to the external algebra $ \Im^{^B} $ rather 
than on the use of the intrinsic field algebra $ \Im $ which defines the model.

\section{Gauge Invariant Formulation}

In the Lagrangian formalism and within the Faddeev-Satashvili
proposal [3], the so-called $GI$
formulation of the anomalous gauge theory is constructed introducing
extra degrees of freedom into the theory by adding to the original $GNI$ 
Lagrangian a WZ term [3,4]
$$ {\cal L}_{_{_{GI}}}\,=\,{\cal L}_{_{_{GNI}}}\,+\,{\cal L}_{_{_{WZ}}}
\,\,, \eqno(3.1) $$
where $ {\cal L}_{_{_{WZ}}} $ is the WZ Lagrangian density given by
$$ {\cal L}_{_{_{WZ}}}\,=\,\frac{1}{2}\,(\,a\,-\,1\,)\,(\,
{\partial }_{\mu }\,\theta \,)^2\,+\,\frac{g}{\sqrt \pi }\,{\cal A}_{\,\mu }\,\lbrace
\,(\,a\,-\,1\,)\,{\partial }^{\,\mu }\theta \,-\,{\tilde {\partial }}^{\,\mu }
\theta\,\rbrace \,\,\,.\eqno (3.2) $$
The resulting ``embedded' theory exhibits invariance under the {\it extended} local gauge transformations
$$ \psi (x)\,\rightarrow\,e^{\,i\,\Lambda (x)\,P_+}\,\psi (x)\,\,,
\eqno(3.3a) $$
$$ {\cal A}_\mu (x)\,\rightarrow\,{\cal A}_\mu (x)\,+\,\frac{1}{g}\,\partial
_\mu\, \Lambda (x)\,\,, \eqno(3.3b) $$
$$ \theta (x)\,\rightarrow \,\theta (x)\,+\,\Lambda (x)\,\,. \eqno(3.3c) $$

Within the path-integral and operator approaches, the $GI$ formulation
is constructed by enlarging the intrinsic local field algebra $
\Im_{_{_{GNI}}} $ through an operator-valued gauge transformation on 
the $GNI$ formulation of the anomalous model [5,6,7],
$$ ^\theta\!\psi (x)\,=\,:\,e^{\,-\,i\,\sqrt\pi\,\theta
(x)\,P_+}\,\psi (x)\,:\,\,, \eqno(3.4a) $$
$$ ^\theta\!\!{\cal A}_\mu (x)\,=\,{\cal A}_\mu (x)\,-\,\frac{\sqrt \pi}{g}\,
\partial_\mu\, \theta (x)\,\,. \eqno(3.4b) $$

The gauge-transformed Lagrangian density is given by [7]
$$ {\cal L}_{_{GI}}\,=\,-\,\frac{1}{4}\,(\,{\cal F}_{\mu \nu }\,)^{\,2}\,
+\,i\,^\theta\!\bar{\psi }\,\partial\!\!\!/\,^\theta\!\psi \,+\,
\frac{g}{\sqrt \pi}\,^\theta{\cal K}_{_{\ell}}^{\,\mu}\,\,^\theta\!\!{\cal A}_{\mu }
\,+\,\frac{a g^{\,2}}{2\pi}
\,(\,^\theta\!\!{\cal A}_{\mu })^{\,2}\,\,,
\eqno (3.5)$$
which can be written as
$$ {\cal L}_{_{GI}}\{\,^\theta\!\bar\psi, \,^\theta\!\psi,
\,^\theta\!\!{\cal A}_\mu\,\}\,=\,{\cal L}_{_{_{GNI}}}\{\,\bar\psi, \psi,
{\cal A}_\mu\,\}\,+\,{\cal L}_{_{_{WZ}}}\{\,{\cal A}_\mu, \theta\,\}\,\,.
\eqno(3.6) $$
The bosonized effective $GI$ Lagrangian density is given by (2.5) plus
the WZ Lagrangian (3.2).

The set of gauge-transformed field operators $ \{^\theta\!\bar \psi,\,^\theta\!\psi,\,
^\theta\!\!{\cal A}_\mu \} $ is
invariant (by construction) under the extended local gauge 
transformations (3.3) and generates the $GI$ field 
algebra $ ^\theta \Im \,\equiv\,\Im_{_{_{GI}}} $.

Within the formulation of the theory based on the 
local field algebra generated from the intrinsic irreducible set of
field operators $ \{ \bar \psi, \psi, {\cal A}^\mu \} $, the
procedure of defining the $GI$ version of the anomalous theory make use
of the introduction of a Bose field algebra, generated by the WZ
field $ \theta $, which is an external algebra to the intrinsic field
algebra $ \Im_{_{_{GNI}}} $. We  shall
denote 
by $\,^\theta{\cal H}\,\equiv\,{\cal H}_{_{_{GI}}}\,\doteq\,^\theta \Im\,
\Psi_o $, the state space 
on which the field algebra generated from the set of
gauge-transformed field operators $ \{\,^\theta\!\bar \psi,\,
^\theta\!\psi,\,^\theta\!\!{\cal A}_\mu\,\} $ is represented.

The question that arises refers to whether the introduction of the external
WZ field algebra changes the physical content of the original
anomalous field theory. Without neglecting the required mathematical
rigor, the algebraic isomorphism between the $GNI$ and $GI$
formulations of the anomalous chiral model can be established
starting from the enlargement of the intrinsic field algebra $
\Im_{_{_{GNI}}} $. As we will show the introduction of
the WZ field enlarges the theory and replicates it, without changing
either its algebraic structure or its physical content. As a matter of 
fact, the 
implementation of the operator-valued gauge transformation connecting the $GNI$ 
and $GI$ formulations is undercover of the algebraic isomorphism
between the corresponding field algebras. This is most easily
understood in the bosonization approach in which the field 
algebras $ \Im^{^B}_{_{_{GNI}}} $ and $ \Im_{_{_{GNI}}} $ are available.

To begin with and following along the same lines as those of
Refs.[8,9], we consider the enlargement of the Bose field 
algebra $ \Im ^{^B}_{_{_{GNI}}} $ by the introduction of the WZ field
through 
$$ \theta\,=\,\xi^{\prime}\,-\,\frac{g}{\sqrt \pi}\,\Bigl (\,\frac{1}{a-1}\,
\chi\,+\,\lambda\,\Bigr )\,\,, \eqno(3.7) $$
which corresponds to shift the field $ \lambda ^\prime $ by
$$ -\,\frac{g}{\sqrt \pi }\,\lambda ^\prime \,=\,\theta \,-\,\xi
^\prime \,\,,\eqno (3.8) $$
with the fields satisfying the algebraic constraints
$$ [\,\lambda^\prime (x)\,,\,\xi ^\prime (y)\,]\,=\,0\,\,,\eqno (3.9a) $$
$$ [\,\theta (x)\,,\,\xi ^\prime (y)\,]\,=\,-\,\frac{g}{\sqrt \pi}\,
[\,\theta (x)\,,\,\lambda ^\prime (y)\,]\,=\,[\,\xi ^\prime (x)\,,\,
\xi ^\prime (y)\,]\,=\,
\frac{g^2}{\pi}\,[\,\lambda ^\prime (x)\,,\,\lambda ^\prime
(y)\,]\,\,. \eqno (3.9b) $$
As we shall see, the shift (3.8) together with the commutations relations 
above leads to a gauge invariant algebra and ensures that no additional 
physical degree of freedom is introduced into the theory.

With the field re-definition (3.8) we
introduce a Bose field subalgebra $ \Im^{^B}_{_{_{GI}}}\,
\subset\,\Im^{^B}_{_{_{GNI}}} $, generated by the set of building 
blocks $ \{\,\phi^{\prime}, \chi_{_2},\chi_{_1}, \xi^{\prime}\,\} $, and the 
Hilbert space $ {\cal H}^{^B}_{_{_{GNI}}}\,
\doteq\,\Im_{_{_{GNI}}}^{^B}\,\Psi_o $ can be decomposed as
$$ {\cal H}^{^B}_{_{_{GNI}}}\,=\,{\cal H}_{_{\chi_{_1}}}\otimes
{\cal H}_{_{\phi^\prime}}\otimes{\cal H}_{_{\chi_{_2}}}\otimes
{\cal H}_{_{\lambda^\prime}}\,=\,
{\cal H}_{_{\chi_{_1}}}\otimes{\cal H}_{_{\phi^\prime}}\otimes 
{\cal H}_{_{\chi_{_2}}}\otimes{\cal H}_{_{\xi^\prime}}\otimes 
{\cal H}_{_{\theta}}\,\equiv\,
{\cal H}^{^B}_{_{_{GI}}}\otimes {\cal H}_{_{\theta}}\,\,, \eqno (3.10) $$
where the subspace $ {\cal H}^{^B}_{_{_{GI}}}\,\subset\,
{\cal H}^{^B}_{_{_{GNI}}} $ is defined 
by $ {\cal H}^{^B}_{_{_{GI}}}\,\doteq \,\Im^{^B}_{_{_{GI}}}\Psi_o $. 

Using (3.8), the set of 
field 
operators $ \{\bar \psi_{_{_{GNI}}}, \psi _{_{_{GNI}}}, {\cal A}^\mu _{_{_{GNI}}} \} $ 
 given by Eq.(2.19), and which defines the $GNI$ algebra $ \Im _{_{_{GNI}}} $, can be written as
$$ \psi _{_{_{GNI}}}(x)\,=\,:\,\psi_{_{_{GI}}}(x)\,e^{\,
 i \sqrt \pi\,P_+\,\theta (x)}\,:\,=\,{\cal Z}^{1 / 2}\,\psi_{_{_{GI}}}(x)\,:\,e^{\,
 i \sqrt \pi\,P_+\,\theta (x)}\,:\,, \eqno (3.11)$$
$$ {\cal A}^\mu_{_{_{GNI}}}(x)\,=\,{\cal A}^\mu_{_{_{GI}}}(x)\,+\,
\frac{\sqrt \pi}{g}\,\partial ^\mu\,\theta (x)\,\,, \eqno (3.12) $$
where 
$$ {\cal Z}\,=\,\exp \{\,4 \pi\,\langle\,\theta (x + \epsilon)\,\xi
^\prime (x)\,\rangle_o\,\}\,=\,[\,- \mu \,\epsilon ^2\,]^{1/(a -
1)}\,\,\eqno (3.13) $$
is a finite wave function renormalization constant ($a > 1$), and the
longitudinal current can be written as
$$ L^\mu_{_{_{GNI}}}(x)\,=\,L^\mu_{_{_{GI}}}(x)\,-\,{\cal J}^\mu_{_{_{WZ}}}(x)\,\,,
\eqno(3.14) $$
where the WZ current is
$${\cal J}^{\mu}_{_{_{WZ}}}\,=\,-\,\frac{g}{\sqrt \pi }\,\lbrace
\,(\,a\,-\,1\,)\,{\partial }^{\,\mu }\theta \,-\,{\tilde {\partial }}^{\,\mu }
\theta\,\rbrace \,\,\,,\eqno (3.15) $$
and
$$  \langle \,L^{\mu }_{_{_{GI}}} \Psi_o\,,\,L^{\nu }_{_{_{GI}}} \Psi _o
\,\rangle\,=\,0\,\,. \eqno (3.16) $$

The anomalous 
nature of the model induces the transformation (3.14) for the longitudinal
piece of the current [7], such that the Hilbert 
space $ {\cal H}_{_{_{GI}}}\,\doteq
\,\Im _{_{_{GI}}} \Psi_o $ is defined by the subsidiary condition
$$\langle \,\Phi\,,\,(\,\,{\cal J}_{_{_{GI}}}^{\,\nu}(x)\,-\,\partial _{\mu}\,
{\cal F}^{\,\mu \nu}(x)\,)\,\Psi \,\rangle \,=\,\langle \,\Phi\,,\,
L_{_{_{GI}}}^{\,\nu }(x)\,\Psi\,\rangle\,=\,0\,\,\,\,\,,
\,\,\,\,\,\forall\, \Phi \,,\, \Psi \,\,\in\,\,
{\cal H}_{_{_{GI}}}\,\,, \eqno (3.17) $$
with
$$ ^\theta {\cal J}^\mu\,\equiv\,{\cal J}^\mu_{_{_{GI}}}(x)\,=\,{\cal J}^\mu_{_{_{GNI}}}(x)\,+\,
{\cal J}^\mu_{_{_{WZ}}}(x)\,\,.
\eqno(3.18) $$

Using (3.8) together with (2.17b), yields
$$  \phi _{_{\ell}}\,-\,\theta\,=\,\frac{1}{2} (\phi - \tilde \phi ) =\,
\phi _{_{\ell}}^{\,\prime}\,+ \,
\frac{1}{\sqrt {a - 1}}\,( \chi _{_2} - \chi _{_1} )\,-\,\xi^{\prime}\,\,.
\eqno (3.19)$$
Note that the expression for the field 
combination $ \phi _{_{\ell}} - \theta $ is the same 
as that for $  \phi _{_{\ell}} $ given by eq.(2.17b), but with
the non canonical free field $ \lambda^{\prime} $ replaced 
by $ \xi^{\prime} $. In agreement with one of the main conclusions of
Ref.[8], we find that {\it the field algebra $ \Im_{_{_{GNI}}}^{^B} $ of the $GNI$ formulation
generated from the building 
blocks $ \{\,\phi^{\prime}, \chi_{_2},\chi_{_1}, \lambda^{\prime}\,\} $ is 
replaced in the $GI$ formulation by the 
algebra $ \,\Im_{_{_{GI}}}^{^B} $ generated from the set of 
fields $ \{\,\phi^{\prime}, \chi_{_2},\chi_{_1}, \xi^{\prime}\,\} $}. 

Using the transformations (2.10) and (3.8), we obtain the bosonized
version of the $GI$ Lagrangian as given by [8]
$$ {\cal L}_{_{GI}}\,=\,\frac{1}{2}\,(\,\partial _{\mu}\,\phi ^{\,\prime}\,)^{\,2}
\,-\,\frac{1}{2}\,(\,\partial _{\mu}\,\chi _{_2}\,)^{\,2}\,+\,
\frac{(a\,-\,1)}{2}\,(\,\partial _{\mu}
\,\xi ^{\,\prime}\,)^{\,2}\,+\,\frac{1}{2}\,(\,\partial _{\mu}\,\chi_{_1}\,)^{\,2}\,
-\,\frac{1}{2}\,m_a^{\,2}\,\chi _{_1}^{\,2}\,\,. \eqno (3.20) $$

The \, formal \, expressions \,
for \, the\, operators\, $ \{\,\psi_{_{_{GI}}},\,{\cal A}^\mu_{_{_{GI}}},\,
L^\mu_{_{_{GI}}}\,\} $ \,\,are \,\,the \,\,same \,\,as \hfill{for} $ \{\,\psi_{_{_{GNI}}},\,
{\cal A}^\mu_{_{_{GNI}}},\,L^\mu_{_{_{GNI}}}\,\} $, except for the
replacement of the 
field $ g/\sqrt \pi \lambda ^\prime $ by the field $ \xi ^\prime $. The 
set 
of field operators $ \{ \bar\psi _{_{_{GI}}},\,\psi _{_{_{GI}}},\,
{\cal A}^\mu _{_{_{GI}}} \} $ defines (through polynomials of these 
smeared fields, Wick ordering, point-splitting regularization
of polynomials, etc.) the $GI$ algebra $ \Im_{_{_{GI}}} $ which is subject 
to the constraint
$$ [\,{\cal O}\,,\,L^{\mu }_{_{_{GI}}}\,]\,=\,0\,\,\,,\,\forall\,{\cal O}\,\in
\,\Im_{_{_{GI}}} \,\,\,.\eqno (3.21) $$

Although the so-introduced WZ field has acquired dynamics, it is a redundant 
field in the sense that it does not change the algebraic structure of
the model and therefore does not change its physical content. As we
shall see, this
implies the isomorphism between the field algebras $ \Im_{_{_{GNI}}}
$ and $ \Im_{_{_{GI}}} $. 

From the algebraic point of view, the WZ field introduced according
with (3.8) and subjected to the commutation relations (3.9), plays a 
redundant role on the structure of the Bose field
algebra $ {\cal H}^{^B}_{_{_{GNI}}} $. This can be seen by considering
the Wightman functions associated with the Wick exponential of the 
field $ \lambda ^\prime $:
$$ \langle\,\prod_{j = 1}^n\,:e^{\,-i\,g\,\lambda ^\prime
(x_j)}:\,\prod_{k =
1}^n\,:e^{\,i\,g\,\lambda^\prime(y_k)}:\,\rangle\,=\,
\langle\,\prod_{j = 1}^n\,:e^{\,i\,\sqrt \pi\,[\,\theta (x_j)\,-\,
\xi ^\prime
(x_j)\,]}:\,\prod_{k = 1}^n\,:e^{\,-i\,\sqrt \pi\,[\,\theta (y_k)\,-
\xi^\prime(y_k)\,]}:\,\rangle\,= $$
$$ =\, 1\,\times\,
\langle\,\prod_{j = 1}^n\,:e^{\,-\,i\,\sqrt \pi\,\xi ^\prime
(x_j)}:\,\prod_{k = 1}^n\,:e^{\,i\,\sqrt \pi
\,\xi^\prime(y_k)}:\,\rangle\,\equiv $$
$$
\equiv\,\langle\,\prod_{j = 1}^n\,:e^{\,-\,i\,[\,g\,\lambda ^\prime (x_j)\,-\,
 \sqrt \pi\,\theta (x_j)\,]}:\,\prod_{k =
1}^n\,:e^{\,i\,[\,g\,\lambda^\prime(y_k)\,-\,\sqrt \pi\,\theta (y_k)\,]}:\,
\rangle \,\,,\eqno (3.22) $$
in which we have explicited the identity arising from the commutator factors
$$ 1\,=\,\exp \,\Bigl \{\,\pi\sum_{j,\,k\,=\,1}^n\,\Bigl (\,
\langle\,\theta (x_j)\,\theta (y_k)\,
\rangle _o\,-\,\langle\,\theta (x_j)\,{\xi^\prime} (y_k)\,
\rangle_o\,-\,\langle \,{\xi^\prime} (x_j)\,\theta
(y_k)\,\rangle_o \,\Bigr )\,+ \Bigr.$$
$$+\,\pi\sum_{i < j}^n\,\Bigl (\,
\langle\,\xi^\prime (x_i)\,\theta (x_j)\,
\rangle _o\,+\,\langle\,\theta (x_i)\,{\xi^\prime} (x_j)\,
\rangle_o\,-\,\langle \,\theta (x_i)\,\theta (x_j)\,\rangle_o
\,\Bigr )\,+ $$
$$\Bigl. + \,\pi\sum_{k < l}^n\,\Bigl (\,
\langle\,\xi^\prime (y_k)\,\theta (y_l)\,
\rangle _o\,+\,\langle\,\theta (y_k)\,{\xi^\prime} (y_l)\,
\rangle_o\,-\,\langle \,\theta (y_k)\,\theta (y_l)\,\rangle_o
\,\Bigr )\,\Bigr \}\,\,.\eqno (3.23)$$
As a consequence of the identity (3.22) we obtain
$$ \langle\,\prod_{j = 1}^n\,\psi^\dagger_{_{_{GNI}}}(x_i)\,\prod_{k
= 1}^n\,\psi_{_{_{GNI}}}(y_k)\,\rangle\,\equiv\,
 \langle\,\prod_{j = 1}^n\,\psi^\dagger_{_{_{GI}}}(x_i)\,\prod_{k
= 1}^n\,\psi_{_{_{GI}}}(y_k)\,\rangle \,\,,\eqno (3.24) $$
expressing the isomorphism between the Wightman functions generated
from the field operator $ \psi _{_{_{GNI}}}\,\in\,\Im_{_{_{GNI}}} $ and 
those generated from the field operator $ \psi _{_{_{GI}}}\,\in\,
\Im_{_{_{GI}}} $. In the same way, we get
$$ \langle\,\prod_{j = 1}^n\,{\cal A}^\mu_{_{_{GNI}}}(x_i)\,\prod_{k
= 1}^n\,{\cal A}^\nu_{_{_{GNI}}}(y_k)\,\rangle\,\equiv\,
 \langle\,\prod_{j = 1}^n\,{\cal A}^\mu_{_{_{GI}}}(x_i)\,\prod_{k
= 1}^n\,{\cal A}^\nu_{_{_{GI}}}(y_k)\,\rangle \,\,.\eqno (3.25) $$

From the above considerations, the 
isomorphism between the field algebra $ \Im_{_{_{GNI}}} $ defining the $GNI$
formulation and the algebra $ \Im_{_{_{GI}}} $ defining the $GI$
formulation of the CSM implies that the state 
space $ {\cal H}_{_{_{GNI}}} $, which provides a representation of the $GNI$ intrinsic
local field algebra $ \Im_{_{_{GNI}}} $, is isomorphic to the state
space $ {\cal H}_{_{_{GI}}} $ on which the $GI$ local field algebra is 
represented; i.e.,
$$ \langle\,\wp \,\{\,\bar\psi_{_{_{GNI}}},\,\psi_{_{_{GNI}}},\,
{\cal A}^\nu_{_{_{GNI}}}\,\}\,\rangle\,\equiv\,
\langle\,\wp \,\{\,\bar\psi_{_{_{GI}}},\,\psi_{_{_{GI}}},\,
{\cal A}^\nu_{_{_{GI}}}\,\}\,\rangle\,\equiv \langle\,
\wp \,\{\,^\theta\!\bar\psi,\,^\theta\!\psi,\,^\theta\!\!{\cal A}^\nu\,\}\,
\rangle\,\,, \eqno (3.26) $$
where {\Large $ \wp $} is any polynomial in the intrinsic field
operators. 

The fact that the Wightman functions of the $GNI$
formulation are the same of those of $GI$ formulation is in agreement
with the results of Refs.[6,8] in the sense that {\it the introduction of
the WZ field enlarges the theory and replicates it, without changing
neither its algebraic structure nor its physical content}. Consequently 
the same analysis made in
section 2 for the $GNI$ formulation applies to its isomorphic $GI$ 
formulation. The conclusions referring to cluster 
decomposition and $ \Theta $-vacuum parametrization and the suggested
equivalence of VSM and CSM defined for $ a = 2 $, are the same for the $GI$ 
formulation. Therefore we completely disagree with the conclusion
of the authors of Refs.[9,10] who claim the need for a $ \Theta
$-vacuum parametrization and the equivalence of the VSM and the $GI$
formulation of CSM defined for $ a = 2 $.

Now, let us make some final remarks on the suggested connection with the 
VSM. In Ref.[9] it is claimed that for $ a = 2 $ and apart from free
physical chiral states, the $GI$ formulation of CSM is equivalent to the 
VSM. These statements disagree with our conclusions \footnote{\it And also
disagrees with the conclusions of Refs.[6,8] and [12].} about the behavior 
of the fermionic particle in both $GNI$ and $GI$ formulations and are
misleading in view of the algebraic isomorphism between $GNI$ and $GI$ 
formulations. 

The local charge operator $ {\cal Q}_{_{_{\varphi}}}\,=\,\lim_{R
\rightarrow \infty} {\cal Q}_{_{_{\varphi,R}}} $, associated with a field
operator $ \varphi $, defines automorphisms 
of the field algebras $ ^{^B}\Im^{^{ext}}_{_{_{\alpha}}} $, $ \Im^{^B}_{_{_{\alpha}}} $
and $ \Im_{_{_{\alpha}}} $, which are implementable in the various 
spaces $ ^{^B}{\cal H}^{^{ext}}_{_{_{\alpha}}} $, $ {\cal H}^{^B}_{_{_{\alpha}}} $
and $ {\cal H}_{_{_{\alpha}}} $. Using the algebraic constraints (3.9), we can
display the following implementability  conditions 
$${\cal Q}_{_{_{\lambda^\prime}}}{\cal H}^{^B}_{_{_{GNI}}}\,\neq\,0\,\,\,,
\,\,\,{\cal Q}_{_{_{\lambda^\prime}}}{\cal H}_{_{_{GNI}}}\,\neq\,0\,\,\,, $$
$${\cal Q}_{_{_{\lambda^\prime}}}{\cal H}^{^B}_{_{_{GI}}}\,=\,0\,\,\,,
\,\,\,{\cal Q}_{_{_{\lambda^\prime}}}{\cal
H}_{_{_{GI}}}\,=\,0\,\,\,;\eqno (3.27a)$$
$${\cal Q}_{_{_{\xi^\prime}}}{\cal H}^{^B}_{_{_{GNI}}}\,=\,0\,\,\,,
\,\,\,{\cal Q}_{_{_{\xi^\prime}}}{\cal H}_{_{_{GNI}}}\,=\,0\,\,\,,$$
$${\cal Q}_{_{_{\xi^\prime}}}{\cal H}^{^B}_{_{_{GI}}}\,\neq\,0\,\,\,,
\,\,\,{\cal Q}_{_{_{\xi^\prime}}}{\cal H}_{_{_{GI}}}\,\neq\,0\,\,\,;
\eqno (3.27b)$$
and 
since $ \theta\,=\,\xi^\prime\,-\,\frac{g}{\sqrt\pi}\,\lambda^\prime $, we get
$${\cal Q}_{_{_{\theta}}}{\cal H}^{^B}_{_{_{GNI}}}\,\neq\,0\,\,\,,
\,\,\,{\cal Q}_{_{_{\theta}}}{\cal H}_{_{_{GNI}}}\,\neq\,0\,\,\,,$$
$${\cal Q}_{_{_{\theta}}}{\cal H}^{^B}_{_{_{GI}}}\,\neq\,0\,\,\,,
\,\,\,{\cal Q}_{_{_{\theta}}}{\cal H}_{_{_{GI}}}\,\neq\,0\,\,\,.\eqno (3.27c) $$
The latter
condition implies that the Wick exponential and derivatives of the WZ
field $ \theta $ can be defined on both Hilbert 
spaces $ {\cal H}_{_{_{GNI}}} $ and $ {\cal H}_{_{_{GI}}} $, as
implied by the operator-valued gauge transformation (3.4) which
enables the relationship between the isomorphic $GNI$ and $GI$ formulations.

Following along the same lines as those of section 2.4 for the $GNI$
formulation, for $ a = 2 $ the Fermi field 
operator $ {\psi_{_{\!\!\ell}}}_{_{_{GNI}}} $ can be 
written as
$$ {\psi _{_{\!\!\ell}}}_{_{_{GNI}}} =\,
:\,{\psi^\prime_{_{\!\!\ell}}}_{_{_{GI}}}\,
e^{2\,i\,\sqrt \pi \,\theta_{_{r}}}:\,
:e^{2\,i\,\sqrt \pi\,(\,\theta_{_{\ell}}\,-\,
\xi^\prime_{_{\ell}})}\,:\,\,,\eqno (3.28)  $$
where the expression for the 
operator $ {\psi^\prime_{_{\!\!\ell}}}_{_{_{GI}}} $ is the same of
Eq.(2.47), only replacing the field $ \lambda_{_{r}}^\prime $ by the field $
\xi_{_{r}}^\prime $.

The trivialization of the charge $ {\cal Q}_{_{_{\tilde\lambda^\prime}}} $, carried
by the field combination $ \theta_{_{\ell}}\,-\,\xi^\prime_{_{\ell}}\,=\,
\lambda^\prime_{_{\ell}}\,\in\,^{^B}\Im^{^{ext}}_{_{_{GNI}}} $, in the 
restriction 
from $ ^{^B}{\cal H}^{^{ext}}_{_{_{GNI}}} $ to $ {\cal H}_{_{_{GNI}}} $, implies 
that the
closure of the local states associated to the field algebra $
\Im_{_{_{GNI}}} $ intrinsic to the model does not allow the
introduction of the operator $ : \exp\,\{2\,i\,\sqrt \pi\,
(\,\theta_{_{\ell}}\,-\,\xi^\prime_{_{\ell}})\} : $, which is charged
under $ {\cal Q}_{_{_{\tilde\lambda^\prime}}} $. This implies that the field 
operator (see Eq.(2.47))
$$ S_{_{Sch}}\,\doteq\,:\,e^{\,+\,2\,i\,g\,\lambda_{_{\ell}}^{\prime}}\,
\psi_{_{\!\!\ell}}\,:\,\vert_{_{a=2}}\,\equiv\, 
:e^{-\,2\,i\,\sqrt \pi\,(\,\theta_{_{\ell}}\,-\,
\xi^\prime_{_{\ell}})}:\,{\psi _{_{\!\!\ell}}}_{_{_{GNI}}}\,=\,
:\,{\psi^\prime_{_{\!\!\ell}}}_{_{_{GI}}}\,
e^{2\,i\,\sqrt \pi \,\theta_{_{r}}}:\,\,,\eqno (3.29) $$
cannot be defined on $ {\cal H}_{_{_{GNI}}} $ and does not belong to
the field algebra $ \Im_{_{_{GNI}}} $.

The enlargement of the Boson field algebra $ \Im^{^B}_{_{_{GNI}}} $ performed 
in Ref.[9], with the introduction of the dual WZ field
$$\tilde \theta\,=\,\tilde\xi^{\prime}\,-\,\frac{g}{\sqrt \pi}\,\Bigl (\,\frac{1}{a-1}\,
\chi\,+\,\lambda\,\Bigr )\,\,,\,\forall\,a\,>\,1\,\,, \eqno(3.30) $$
implies that
$$ \theta_{_{_{\ell}}}\,=\,\xi^\prime_{_{_{\ell}}}\,\,,\eqno (3.31) $$
and thus $ \lambda^\prime_{_{_{\ell}}}\,\equiv\,0 $. In this case
the field operator (3.30) reduces to
$$ {\psi _{_{\!\!\ell}}}_{_{_{GNI}}} =\,
:\,{\psi^\prime_{_{\!\!\ell}}}_{_{_{GI}}}\,e^{2\,i\,\sqrt \pi\,\theta_{_{_{r}}}
}\,:\,\,.\eqno (3.32)  $$

For the moment consider the following commutation relations 
$$ \bigl [ \phi _{_{\ell}} (x) - \theta (x)\,,\,L_{_{_{GI}}} (y) \bigr ] = 0  \,\,,
\eqno (3.33a)$$
$$  \bigl [ \phi _{_{\ell}} (x) - \theta_{_{\!r}} (x)\,,\,
L_{_{_{GI}}} (y) \bigr ] = \bigl [ \theta_{_{\!\ell}} (x)\,,\,
L_{_{_{GI}}} (y) \bigr ] = \Biggl ( \frac{a - 2}{a - 1} \Biggr ) 
\Delta ( x^{\,+} - y^{\,+} ; 0 ) \,\,, \eqno (3.33b)$$
where $ L_{_{_{GI}}} $ is the potential for the longitudinal current
$ L^\mu_{_{_{GI}}} $. From these commutation relations we see that for values of the 
regularization parameter in the range $ a> 1 $, the field combination
$ \phi_{_{\ell}}\,-\,\theta $ satisfies the subsidiary 
condition (3.21). However, for
the special value $ a = 2 $, there is a broader class of operators belonging
to $ \,^{^B}\Im_{_{_{GI}}}^{^{ext}} $ that satisfy the subsidiary condition and
include the field combination $ \phi_{_{\ell}}\,-\,\theta_{_{r}} $ and the 
field $ \theta_{_{\!\ell}}\,=\,\xi^{\prime}_{_{\ell}} $,  which in this case 
decouples from the longitudinal current (3.14b) but not from the Lagrangian
density, as occurs in the VSM.

However, the use of the external Bose 
algebra $ ^{^B}\Im^{^{ext}}_{_{_{GNI}}} $ makes the problem more
dramatic since the trivialization of the charge $ {\cal
Q}_{_{_{\tilde \theta}}} $, associated with the WZ field 
component $ \theta_{_{r}} $, implies that the Wick 
exponential $ :\exp \{2\,i\,\sqrt \pi\,\theta_{_{r}}\}: $ cannot be
defined either on the Hilbert space $ {\cal H}_{_{_{GNI}}} $ or 
on $ {\cal H}_{_{_{GI}}} $. In this
case the definition of the 
operator $ {\psi^\prime_{_{\!\!\ell}}}_{_{_{GI}}} $ is meaningless as
well\footnote{\it Since for $ a = 2 $ the field 
combination $ (\,\phi_{_{\ell}}\,-\,\theta_{_{\!r}}\,)
\,\in\,^{^B}\Im^{^{ext}} $ satisfies the subsidiary condition, the authors
of Ref.[9] chose the field operator $ ^{\theta_{_{\!r}}}\psi_{_{\!\!\ell}}\,
\not\in\,^{\theta}\Im $ instead 
of $ ^\theta\!\psi_{_{\!\!\ell}}\,\in\,\Im_{_{_{GI}}} $ to define the physical fermionic content of
the model. The gauge transformed operator (2.50) is given by

$$^\theta\Psi (x)\,=\,\Bigl (\,\frac{\mu_o}{2\pi}\,\Bigr )^{\,1/4}\,:\,
e^{\,i\,\sqrt\pi\,\gamma^{\,5}\,\chi_{_1}(x)}\,:\,e^{\,i\,\sqrt\pi\,[\,
\gamma^{\,5}\,^\theta\!\tilde L(x)\,+\,^\theta\!L(x)\,]}\,:\,\,,  $$

\noindent and does not belong to $ ^\theta\Im $ either. However, the fact that the 
operator $ ^\theta\Psi $ of the $GI$ formulation ( or $ \Psi $ of the
$GNI$ formulation ) shares the same features of the fermion operator of
the covariant solution of the VSM [13] does not imply the equivalence
between the two models. In order for  this to be true, such an equivalence should be
established between all Wightman functions of the two models and this
is not the case, since neither $ \Psi\,\not\in\,\Im_{_{_{GNI}}} $ can be defined on 
the Hilbert space $ {\cal H}_{_{_{GNI}}} $ of the $GNI$ formulation nor $
\,^\theta\Psi \,\not\in \,\Im_{_{_{GI}}}  $ can be defined 
in the Hilbert space $ \,{\cal H}_{_{_{GI}}} $ of the $GI$ formulation.

The Hilbert space $ \,^\theta {\cal H}\,\equiv\,{\cal H}_{_{_{GI}}} $ can 
be factorized as a product

$$ ^\theta {\cal H} =  {\cal H}_{\chi_1} \otimes {\cal H}_{{\chi_2},
\xi^{\prime},\psi_o}\,\,. $$

\noindent The Hilbert space completion $ {\cal H}_{{\chi_2},\xi^{\prime},\psi_o} $ 
of the local states of the $GI$ formulation is isomorphic to the
closure of the space of the $GNI$ 
formulation $ {\cal H}_{{\chi_2},\lambda^{\prime},\psi_o} $, and also
cannot be further decomposed.}.

At a glance, the isomorphism between $GNI$ and $GI$ Wightman functions
points towards on behalf of
gauge invariance such that the rhs of the identity (3.26)
exhausts all possible observable content of the theory. As a matter of 
fact, the isomorphism 
established by Eqs.(3.26) has an algebraic
interpretation\footnote{\it As
stressed in Ref.[6b] within the path-integral formalism, this
isomorphism only is implemented on full quantum level where the integration
over $ {\cal A}_\mu $ has also been performed. The isomorphism
expressed by the identity (3.26) does not exist on the level where $
{\cal A}_\mu $ is an external field.} and in
order to
be correctly understood must be seen on the 
basis of the Wightman reconstruction theorem in the sense that the set of Wightman 
functions $ \langle\,
{\mbox{\Large $\wp$} } \,\{\,\bar\psi_{_{_{GNI}}},\,\psi_{_{_{GNI}}},\,
{\cal A}^\nu_{_{_{GNI}}}\,\}\,\rangle $ ($ \langle\,{\mbox{\Large $\wp$}}\,
\{\,\bar\psi_{_{_{GI}}},\,\psi_{_{_{GI}}},\,{\cal A}^\nu_{_{_{GI}}}\,\}\,
\rangle $) uniquely determine (up to isomorphisms) a representation of
the field algebra $ \Im_{_{_{GNI}}} $ ($ \Im_{_{_{GI}}} $). 

For the $GNI$ and $GI$ formulations we have the realization
of the corresponding weak local Gauss's law
$$\langle \,\Phi_{_{_{\alpha}}}\,,\,(\,\,{\cal J}_{_{_{\alpha}}}^{\,0}(x)\,-\,
\partial _{1}\,
{\cal F}^{\,1 0}(x)\,)\,\Psi_{_{_{\alpha}}} \,\rangle \,=\,\langle \,
\Phi_{_{_{\alpha}}}\,,\,L_{_{_{\alpha}}}^{\,0}(x)\,\Psi_{_{_{\alpha}}}\,\rangle\,=\,0\,\,
\eqno (3.34) $$
where $ \alpha $ means that we are working in $GI$ or $GNI$ formulations. We can 
construct the local charges
$$ {\cal Q}_{_{_{\alpha}}} [\Lambda]\,=\,\int_{_{_{z^o}}} \Lambda
(z)\,\Bigl (\,
{\cal J}_{_{_{\alpha}}}^{0}(z)\,-\,\partial _{1}\,
{\cal F}^{\,1\,0}(z)\,\bigr )\,dz^1\,=\,\int_{_{_{z^0}}}\,
\bigl \{\,L_{_{_{\alpha}}}^0 (z) \,\,\Lambda (z)\,-\,L_{_{_{\alpha}}}^1 (z)\,
\tilde\Lambda (z)\,\bigr \}\,dz^1 \,\,, \eqno(3.35) $$
with $ \Box \,\Lambda \,=\,0 $, and which are implementable 
on the Hilbert space $ {\cal H}_{_{_{\alpha}}} $. The corresponding charge 
generators are obtained as weak limits of the above charges and the unitary 
operators 
$$ T_{_{_{\alpha}}} [ \Lambda ] = e^{\,-i\,\frac{\sqrt \pi }{g}\,Q_{_{_{\alpha}}}
 [ \Lambda ]} \eqno(3.36) $$
implement the local gauge transformations. Since we are dealing with an 
anomalous theory, the field 
algebra $ \Im _{_{_{GNI}}}\,(\Im _{_{_{GI}}}) $ is a 
singlet under
local gauge transformations generated by the 
operator $ T_{_{_{GNI}}} [ \Lambda ]\,\in\,\Im _{_{_{GNI}}}\,
(T_{_{_{GI}}} [ \Lambda ]\,\in \,\Im _{_{_{GI}}}) $. In view of
Eqs.(3.17-18), we can consider the field algebra $ \Im_{_{_{GI}}} $ as
a field subalgebra of $ \Im_{_{_{GNI}}}\,\supset\,\Im_{_{_{GI}}} $
such that the operator  $ T_{_{_{GNI}}} [ \Lambda ] $ implements the local 
c-number extended gauge transformations:
$$ T_{_{_{GNI}}}\,[ \Lambda ]\,\psi _{_{_{GI}}}(x)\,T_{_{_{GNI}}}^{\,- 1} [ \Lambda ] = 
e^{\,2\,i\,\sqrt \pi\,\Lambda (x)}\, 
\psi _{_{_{GI}}}(x)\,\,, \eqno(3.37a) $$
$$  T_{_{_{GNI}}}\,[ \Lambda ]\,{\cal  A}_{_{_{GI}}}^{\,\mu }(x)\,
T_{_{_{GNI}}}^{\,- 1} [ \Lambda ] = {\cal A}_{_{_{GI}}}^{\,\mu }(x) + 
\frac{\sqrt \pi}{g}\,\partial ^{\,\mu } \Lambda (x)\,\,, \eqno(3.37b) $$
$$  T_{_{_{GNI}}}\,[ \Lambda ]\,\theta (x)\,T_{_{_{GNI}}}^{\,- 1}[ \Lambda ]\,=\,\theta (x)\,
-\,
\Lambda (x)\,\,, \eqno(3.37c) $$
$$ T_{_{_{GNI}}}\,[ \Lambda ]\,\psi _{_{_{GNI}}}(x)\,T_{_{_{GNI}}}^{\,- 1}[ \Lambda ]\,=\,
\psi _{_{_{GNI}}}(x)\,\,, \eqno(3.37d) $$
$$ T_{_{_{GNI}}}\,[ \Lambda ]\,{\cal  A}_{_{_{GNI}}}^{\,\mu }(x)\,
T_{_{_{GNI}}}^{\,- 1}[ \Lambda ]\,=\,
{\cal A}^{\,\mu }_{_{_{GNI}}}(x)\,\,,\eqno(3.37e) $$
and similarly for the operator $ T_{_{_{GI}}}[\Lambda] $ (with $ \Lambda\,
\rightarrow\,-\,\Lambda $ in the analog of the transformation (3.37c)), under which
the field algebra $ \Im_{_{_{GI}}} $ is a singlet.             

From Eq. (3.27c) we see that the generator that is implementable 
in both $GNI$ and $GI$ Hilbert spaces is obtained as weak limit of the
local charge 
operator constructed from the
WZ current $ {\cal J}^\mu_{_{_{WZ}}} $ which, in view of Eq.(3.14), can be
written as
$${\cal Q}_{_{_{WZ}}}\,=\,{\cal Q}_{_{_{GI}}}\,-\,{\cal Q}_{_{_{GNI}}}\,\,. 
\eqno (3.38) $$

In the ``embedded'' formulation of the anomalous model the local gauge 
invariance is recovered at the quantum level, meaning that there exists
an unitary operator 
$$ T_{_{_{WZ}}} [ \Lambda ] = e^{\,-i\,\frac{\sqrt \pi }{g}\,Q_{_{_{WZ}}}
 [ \Lambda ]} \eqno(3.39) $$
which implements the extended local gauge transformations
$$ T_{_{_{WZ}}}\,[ \Lambda ]\,\psi _{_{_{\alpha}}}(x)\,
T_{_{_{WZ}}}^{\,- 1} [ \Lambda ] = 
e^{\,\pm\,2\,i\,\sqrt \pi\,\Lambda (x)}\,\psi _{_{_{\alpha}}}(x)\,\,, \eqno(3.40a) $$
$$  T_{_{_{WZ}}}\,[ \Lambda ]\,{\cal  A}_{_{_{\alpha}}}^{\,\mu }(x)\,
T_{_{_{WZ}}}^{\,- 1} [ \Lambda ] = {\cal A}_{_{_{\alpha}}}^{\,\mu }(x)\,
\pm \,\frac{\sqrt \pi}{g}\,\partial ^{\,\mu } \Lambda (x)\,\,, \eqno(3.40b) $$
$$  T_{_{_{WZ}}}\,[ \Lambda ]\,\theta (x)\,T_{_{_{WZ}}}^{\,- 1}[ \Lambda ]\,=\,
\theta (x)\,+\,2\,\Lambda (x)\,\,, \eqno(3.40c) $$
with $\alpha $ standing for $ GNI $ and $ GI $ field 
operators, respectively. The
local gauge invariance manifests itself by the existence of the 
generator $ {\cal Q}_{_{_{WZ}}} $ under whose action the isomorphic intrinsic
local field algebras $ \Im_{_{_{GNI}}} $ and $ \Im_{_{_{GI}}} $ are not 
singlets.

Contrary to what happens in the VSM [15], the CSM is not a confining
gauge theory and, due to the absence of spurious vacuum raising
operators belonging to the intrinsic field algebra, the 
model does not exhibit a topological vacuum structure.

\section{Conclusions}

A careful analysis of some basic structural properties of the anomalous CSM 
in the isomorphic $GNI$ and $GI$ operator formulations was performed. In our 
treatment we only use the intrinsic field algebra generated from the  
irreducible set of field operators of the theory in order to construct the 
Hilbert space associated with the Wightman functions that define the model. By
relaxing this careful control on the construction of the Hilbert
space of the theory one may carry along a lot of
superfluous states and some misleading conclusions may arise. 

A peculiar feature of the CSM which differs from the vector case
is the fact that the cluster decomposition property is not violated
for Wightman functions that are representations of the field 
algebra $ \Im  $ generated from the irreducible set of field 
operators $ \{\,\bar\psi, \psi, {\cal A}_{\mu}\,\} $. The
need of the $ \Theta $-vacuum parametrization and the equivalence of
the CSM defined for $ a = 2 $ and the VSM, only emerge due to an
improper factorization of the closure of the Hilbert space $ {\cal H} $ which 
defines the representation of the intrinsic field algebra $ \Im $.
The main results of Refs.[9,10] cannot be regarded as structural
properties of the CSM since they are consequences of the use of a
redundant field algebra belonging to the external algebra $ \Im^{^B} $ rather 
than on the use of the intrinsic field algebra $ \Im $ which defines the model.

\vspace{1.5cm}

\underline{{\it Acknowledgments}}: The authors are grateful to Conselho Nacional de Desenvolvimento
Cient\'{\i}fico e Tecnol\'ogico (CNPq - Brasil) and
Coordena\c{c}\~ao de Aperfei\c{c}oamento de Pessoal de N\'{\i}vel
Superior (CAPES- Brasil) for partial financial support. One of us (C.P.N.) would like to thank
the Physics Department of Universidade Federal Fluminense (UFF) for the kind 
hospitality extended to him. We would like to thank R. L. P. G.
Amaral for useful discussions and comments.

\newpage
\centerline{\bf{References}}
\begin{description}
\item [1 - ] R. Jackiw and R. Rajaraman, {\it Phys. Rev. Lett.} {\bf B 54} (1985), 1219.
\item [2  -] In addition the references quoted here, see also
\item [ \,\, -] R. Rajaraman, {\it Phys. Lett.} {\bf{B\,514}} (1985), 305;
\item [ \,\, -] J. Lott and R. Rajaraman, {\it Phys. Lett.} {\bf{B\,165}} (1985), 321;
\item [ \,\, -] A. Niemi and G. Semenoff, {\it Phys. Lett.} {\bf{B\,175}} (1986), 439;
\item [ \,\, -] M. Chanowitz, {\it Phys. Lett.} {\bf{B\,171}} (1986), 280;
\item [ \,\, -] N. K. Falck and G.Kramer, {\it Ann. Phys. (N.Y.)} {\bf{176}} (1987), 330;
\item [ \,\, -] C. Arag\~ao de Carvalho, K. D. Rothe, C. A. Linhares and H. J. Rothe, {\it Phys. Lett.} {\bf{B\,194}}\\
       (1987), 539.
\item [3  -] L. D. Faddeev and L. Shatashvili, {\it Phys. Lett.} 
            {\bf B 176} (1986), 225;
\item [ \,\, -] L. D. Faddeev, {\it Phys. Lett.} {\bf B\,145} (1984), 81.
\item [4 - ] J. Wess and B. Zumino, {\it Phys. Lett.} {\bf B 37} (1971), 95.
\item [5 - ] O. Babelon,F. A. Shaposnik and C. M. Viallet, {\it Phys. Lett.} {\bf{B\,177}}
           (1986), 385;
\item [ \,\, -] K. Harada and I. Tsutsui, {\it Phys. Lett.} {\bf{B\,183}} (1987), 311.
\item [6 -] H. O. Girotti and K. D. Rothe, {\it Int. J. Mod. Phys.} {\bf{A\,12}} 
           (1989), 3041;
\item [\,\, -] E. Abdalla, M. C. Abdalla and K. D. Rothe, ``Non-Perturbative 
               Methods in 2 Dimensional Quantum Field Theory'',
               World Scientific, 1991.        
\item [7 -] L. V. Belvedere, K. D. Rothe, {\it Mod. Phys. Lett.} {\bf A 10}, (1995), 207. 
\item [8 -] D. Boyanovsky, I. Schmidt and M. F. L. Golterman, {\it Ann. Phys. (N.Y.)} {\bf 185} (1988), 111.
\item [9 -] M. Carena and C. E. M. Wagner, {\it Int. Jour. Mod. Phys},
            {\bf A\,6} (1991), 243.
\item [10-] K. Shizuya, {\it Phys. Lett.} {\bf B\,213} (1988), 298;
\item [\,\, -] S. Miyake and K. Shizuya, {\it Phys. Rev.} {\bf D\,36} (1987), 3791;
\item [\,\, -] T. Berger, N. K. Falck and G. Kramer, {\it Int. Jour. Mod.
               Phys.} {\bf A\,4} (1989) 427.
\item [11-] G. Morchio, D. Pierotti and F. Strocchi, {\it Ann. Phys. (N.Y.)}
              {\bf 188} (1988), 217;
\item [\,\, -] F. Strocchi, ``Selected Topics on
               the General Properties of Quantum Field Theory'', 
               Lecture Notes in
               Physics, vol.51, World Scientific Publishing, 1993; see also
\item [\,\, -] A. Z. Capri and R. Ferrari, {\it Nuovo Cimento} {\bf A\,62} (1981), 273.
\item [12-] H. O. Girotti, H. J. Rothe and K. D. Rothe, {\it Phys. Rev.} {\bf D\,34}
            (1986), 592.    
\item [13-] J. H. Lowenstein and J. Swieca, {\it Ann. Phys. (N.Y.)} {\bf 68} (1971), 172;
\item [\,\, -] L. V. Belvedere, J. A. Swieca, K. D. Rothe and B. Schroer,
              {\it Nucl. Phys.} {\bf B\,153} (1979), 112;
\item [\,\, -] L. V. Belvedere, {\it Nucl. Phys.} {\bf B\,276} (1986), 197.              
\item [14-] M. B. Halpern, {\it Phys. Rev.} {\bf D\,13} (1976), 337.
\item [15-] K. D. Rothe and J. Swieca, {\it Phys. Rev.} {\bf D\,15} (1977), 541.
\item [16-] R. F. Streater and A. S. Wightman, ``PCT, Spin, 
            Statistics  and All That'' W. A. Benjamin, Inc,  New York, 1964.
\item [\,\, -] R. Jost, ``Properties of Wightman Functions'' in
               Lectures on Field Theory and the Many-Body Problem, (E.
               R. Caianiello Ed.), Academic Press, New York, 1961;
\item [\,\, -] R. Haag, ``it Local Quantum Physics, Fields, Particles,
               Algebras'', Texts and Monographs in Physics,
               Springer-Verlag, 1992.
\item [17 -] D. Pierotti, {\it Lett. Math. Phys.} {\bf 15} (1988), 219.              
\item [18 -] H. O. Girotti, H. J. Rothe and K. D. Rothe, {\it Phys. Rev.}
                {\bf D\,33} (1986), 514; 
\item [\,\, -] H. O. Girotti, H. J. Rothe and K. D. Rothe, {\it Phys. Rev.} {\bf D\,34} (1986) 592;                
\item [\,\, -]  N. K. Falck and G.Kramer, {\it Ann. Phys. (N.Y.)} {\bf{176}} (1987), 330.
\item [19 -] Wightman, A. S., in M. Levy (ed.), ``Carg\`ese Lectures in 
               Theoretical Physics'' 1964 (1967).
\end{description}

\end{document}